\documentclass[prb,twocolumn,aps,showpacs,preprintnumbers,amsmath,amssymb,groupedaddress]{revtex4}
\usepackage{graphicx}
\usepackage{dcolumn}
\usepackage{bm}

\begin{document}

\preprint{LMH/ATB-1}

\title{Magnetic order and dynamics of the charge-ordered
antiferromagnet La$_{1.5}$Sr$_{0.5}$CoO$_{4}$}

\author{L. M. Helme}
\author{A. T. Boothroyd}
\email{a.boothroyd@physics.ox.ac.uk}
\homepage{http://xray.physics.ox.ac.uk/Boothroyd}
\author{R. Coldea}
\author{D. Prabhakaran}
\affiliation{Department of Physics, Oxford University, Oxford, OX1
3PU, United Kingdom }

\author{C. D. Frost}
\author{D. A. Keen}
\affiliation{ISIS Facility, Rutherford Appleton Laboratory, Chilton,
Didcot, OX11 0QX, United Kingdom}

\author{L. P. Regnault}
\affiliation{INAC-SPSMS-MDM, CEA-Grenoble, 17 rue des Martyrs,
38054, Grenoble Cedex 9, France}

\author{P. G. Freeman}
\author{M. Enderle}
\author{J. Kulda}
\affiliation{Institut Laue-Langevin, BP 156, 38042 Grenoble Cedex 9,
France}

\date{\today}

\begin{abstract}We describe neutron scattering experiments
performed to investigate the magnetic order and dynamics of
half-doped La$_{1.5}$Sr$_{0.5}$CoO$_{4}$. This layered perovskite
exhibits a near-ideal checkerboard pattern of Co$^{2+}$/Co$^{3+}$
charge order at temperatures below $\sim 800$\,K. Magnetic
correlations are observed at temperatures below $\sim 60$\,K but
static magnetic order only becomes established at $31$\,K, a
temperature at which a kink is observed in the susceptibility. On
warming above $31$\,K we observed a change in the magnetic
correlations which we attribute either to a spin canting or to a
change in the proportion of inequivalent magnetic domains. The
magnetic excitation spectrum is dominated by an intense band
extending above a gap of approximately 3\,meV up to a maximum energy
of 16\,meV. A weaker band exists in the energy range 20--30\,meV. We
show that the excitation spectrum is in excellent quantitative
agreement with the predictions of a spin-wave theory generalized to
include the full magnetic degrees of freedom of high-spin Co$^{2+}$
ions in an axially distorted crystal field, coupled by Heisenberg
exchange interactions. The magnetic order is found to be stabilized
by dominant antiferromagnetic Co$^{2+}$--Co$^{2+}$ interactions
acting in a straight line through Co$^{3+}$. No evidence is found
for magnetic scattering from the Co$^{3+}$ ions, supporting the view
that Co$^{3+}$ is in the $S=0$ state in this material.

\end{abstract}

\maketitle

\section{Introduction}

Hole-doped transition-metal oxide antiferromagnets exhibit a range
of intriguing phenomena, including unconventional superconductivity,
metal--insulator transitions, magnetoelectric behavior and extreme
sensitivity to external stimuli. Their more unusual properties are
often found in association with electronically-ordered states with
nano-scale periodicity arising from competition between electronic,
magnetic and structural degrees of freedom. As well as being of
fundamental interest, these highly correlated phases offer the
possibility to tune the macroscopic properties of a material via
selective control of its microscopic states.

At low doping levels, holes introduced into layered antiferromagnet
insulators have a tendency to segregate into stripes, as observed
for example in layered cuprates\cite{Tranquada-Nature-1995},
nickelates\cite{Chen-PRL-1993,Tranquada-PRL-1994,Yamada-PhysicaC-1994}
and, most recently, in layered cobaltates\cite{Cwik-PRL-2009}.
Stripes are a form of complex electronic order in which the
antiferromagnetic structure is modulated by periodic arrays of
hole-rich antiphase boundaries. With increasing doping,
stripe-ordered systems can evolve into metallic or charge-ordered
insulating states depending on the relative importance of the
electron kinetic energy, Coulomb interactions and associated lattice
strain.

At half doping, many transition-metal oxides exhibit an insulating
charge-ordered phase, the stability of which is generally assisted
by cooperative Jahn--Teller distortions. For layered systems the
charge order naturally takes the form of a checkerboard pattern.
This ordering pattern has been reported in the isostructural ``214"
compounds La$_{1/2}$Sr$_{3/2}$MnO$_4$ (Ref.
\onlinecite{Mn-half-doped}), La$_{3/2}$Sr$_{1/2}$CoO$_4$ (Ref.
\onlinecite{Zaliznyak-PRL-2000}) and La$_{3/2}$Sr$_{1/2}$NiO$_4$
(Ref. \onlinecite{Chen-PRL-1993}). Previous investigations have
revealed a number of interesting features in the order these
compounds. For instance, electronic structure calculations for
La$_{1/2}$Sr$_{3/2}$MnO$_4$ suggest that there is almost no charge
separation between the two inequivalent Mn
sites.\cite{No-Mn-charge-order} In the half-doped nickelate, the
checkerboard charge order rearranges itself spontaneously into a
stripe-like phase at low temperatures due to spin--charge
coupling.\cite{Kajimoto-PRB-2003,Freeman-PRB-2002} Finally, in
La$_{3/2}$Sr$_{1/2}$CoO$_4$, there exists the possibility of
spin-state transitions\cite{Moritomo-PRB-1997} due to the near
degeneracy of different terms of the $d$ electron configuration in
octahedrally-coordinated Co$^{3+}$. In short, these canonical
half-doped perovskites with checkerboard order may not be as simple
as they first appear.

In this paper we report the results of neutron scattering
experiments carried out to study the static and dynamic magnetic
properties of the half-doped cobaltate La$_{3/2}$Sr$_{1/2}$CoO$_4$.
The measurements extend over the entire spectrum of cooperative
magnetic excitations throughout the Brillouin zone, which enables us
to characterize the magnetic excitations fully and to quantify the
exchange interactions that stabilize the magnetic ground state. The
work also sought to establish from the excitation spectrum whether
both of the Co sites are magnetically active, or just one. To this
end we carried out a rather detailed analysis of the data in terms
of a generalized spin-wave model that includes the complete set of
spin and orbital degrees of freedom. We find that the measured
magnetic excitation spectrum is in excellent agreement with the
spectrum calculated for the Co$^{2+}$ site alone. There is no
discernible signal in the spectrum from the Co$^{3+}$ site.

\begin{figure}
\begin{center}
\includegraphics
[width=8.5cm,bbllx=0,bblly=0,bburx=318, bbury=204,angle=0,clip=]
{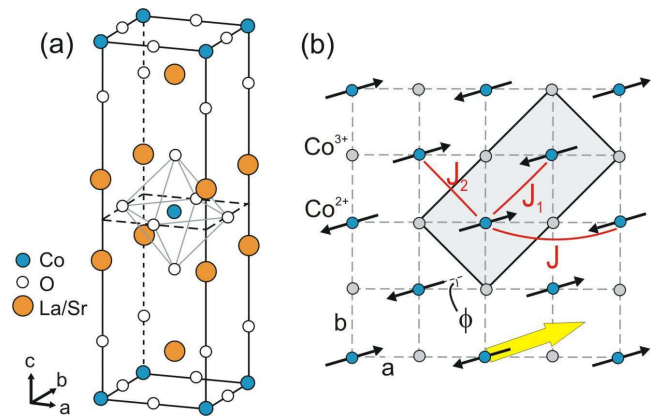} \caption{\label{fig1} (Color online) (a)
Tetragonal unit cell of La$_{2-x}$Sr$_{x}$CoO$_{4}$. (b) Model for
the charge and magnetic order within the $ab$ plane of
La$_{3/2}$Sr$_{1/2}$CoO$_{4}$. The shaded rectangle is the magnetic
unit cell. $\phi$ is the angle of the moments to the $a$ axis, with
positive $\phi$ an anticlockwise rotation. The large (yellow) arrow
represents the projection of the stacking vector ${\bf t} =
(1.5,0.5,0.5)$ for the spin--charge order.}
\end{center}
\end{figure}

The crystal structure of La$_{3/2}$Sr$_{1/2}$CoO$_4$ is described by
the space group $I4/mmm$, with tetragonal unit cell constants
$a=3.84$\,{\AA} and $c=12.5$\,{\AA} [see Fig.\ \ref{fig1}(a)]. The
parent phase La$_2$CoO$_4$ is a Mott insulator which orders
antiferromagnetically below $T_{\rm N}=275$\,K.
\cite{Yamada-PRB-1989} Doping with Sr introduces holes into the
CoO$_2$ layers, and at half doping the 50:50 mixture of Co$^{2+}$
and Co$^{3+}$ ions crystallizes at $T_{\rm co} \simeq 825$\,K into a
checkerboard charge-ordering
pattern,\cite{Zaliznyak-PRL-2000,Zaliznyak-PRB-2001} illustrated in
Fig.\ \ref{fig1}(b). The signature of checkerboard charge order in
neutron diffraction data is a set of peaks at wavevectors ${\bf
Q}_{\rm co}=(h + 0.5,k+ 0.5,l)$, with $h$, $k$ and $l$ integers
(although in the $l$ direction the peaks are very broad). Since
neutrons do not couple directly to charge order, these peaks
actually originate from associated modulations in the shape of the
oxygen octahedra.\cite{Zaliznyak-PRL-2000,Zaliznyak-PRB-2001}

Magnetic order occurs in La$_{3/2}$Sr$_{1/2}$CoO$_4$ at a
temperature well below the charge ordering temperature. Magnetic
susceptibility measurements (Ref \onlinecite{Moritomo-PRB-1997}, see
also Fig.\ \ref{fig4}) revealed a broad maximum in the in-plane
susceptibility at $\sim 60$\,K indicative of a build-up of magnetic
correlations. The data also show a large difference between the
in-plane ($\chi_{ab}$) and out-of-plane ($\chi_{c}$)
susceptibilities, with $\chi_{ab}$ greater than $\chi_{c}$ by at
least a factor two, revealing strong planar anisotropy. Assuming the
Co$^{2+}$ ions are in the high-spin (HS) state with effective spin
$S = 3/2$, as found\cite{Yamada-PRB-1989} in La$_2$CoO$_4$, Moritomo
{\it et al.}\cite{Moritomo-PRB-1997} concluded from the measured
effective moment per Co that the Co$^{3+}$ ions in
La$_{3/2}$Sr$_{1/2}$CoO$_4$ are also in the HS state ($S=2$) and
must therefore carry a moment. However, this analysis did not take
into account the considerable unquenched orbital moment of Co$^{2+}$
in the distorted octahedral field. Two recent studies, one in which
the magnetic susceptibility was analysed with a full atomic
multiplet calculation\cite{Hollmann-NJP-2008} and the other
employing soft x-ray absorption spectroscopy to probe the atomic
levels directly\cite{Chang-PRL-2009}, have confirmed that the
Co$^{2+}$ ions are in the HS state ($S=3/2$) but concluded that the
Co$^{3+}$ ions are in the low-spin (LS) state with $S=0$, as found
for example in LaCoO$_3$.\cite{Raccah-PR-1967} If we accept the
weight of evidence in favor of the LS state then we can assume that
the Co$^{3+}$ ions are non-magnetic apart perhaps from a small Van
Vleck moment induced by the exchange field from Co$^{2+}$ in the
ordered phase.

Neutron diffraction measurements performed by Zaliznyak and
coworkers\cite{Zaliznyak-PRL-2000} confirmed the presence of
magnetic order with a gradual onset starting around $60$\,K. The
magnetic order is characterized by a fourfold group of magnetic
diffraction peaks with slightly incommensurate wavevectors ${\bf
Q}_{\rm m}=(h + 0.5,k+ 0.5,l_1) \pm (0.25-\epsilon
,0.25-\epsilon,0)$ and $(h + 0.5,k+ 0.5,l_2) \pm (-0.25+\epsilon
,0.25-\epsilon,0)$, where $h$ and $k$ are integers, $l_1$ ($l_2$) is
an odd (even) integer, and $\epsilon=0.005-0.008$ depending on
sample preparation.\cite{Zaliznyak-JAP-2004} If the small
incommensurability $\epsilon$ is neglected then the simplest
magnetic structure consistent with the diffraction data would be a
collinear antiferromagnet with ordered moments on the Co$^{2+}$ ions
and propagation vector $(0.25, 0.25)$ within the $ab$ plane, as
shown in Fig.\ \ref{fig1}(b). The in-plane orientation of the
moments ($\phi$) will be addressed later in this paper. It has been
proposed that the observed incommensurability is caused by stacking
faults.\cite{Savici-PRB-2007} The ideal ordering pattern in Fig.\
\ref{fig1}(b) gives rise to magnetic Bragg peaks at in-plane
wavevectors $(h + 0.5,k+ 0.5) \pm (0.25,0.25)$. The observed
fourfold pattern of peaks arises because in tetragonal symmetry a
90\,deg rotation generates an equivalent magnetic structure with
propagation vector $(-0.25, 0.25)$, and in a real sample both
wavevector domains are expected to be present in equal proportion.

A preliminary report of a subset of the data presented here was
given in Ref. \onlinecite{Helme-PhysicaC-2004}, and as far as we are
aware no other measurements of the magnetic excitations in
La$_{3/2}$Sr$_{1/2}$CoO$_4$ have been published. Recently, however,
there appeared a brief account of the magnetic excitation spectrum
of La$_{3/2}$Ca$_{1/2}$CoO$_4$.\cite{Horigane-JMMM-2007} The data
indicate that the energy scale of the magnetic excitations in the
Sr-doped and Ca-doped cobaltates is very similar, but an important
difference is that in La$_{3/2}$Ca$_{1/2}$CoO$_4$ there are ordered
magnetic moments on \emph{both} Co$^{2+}$ the Co$^{3+}$
sites.\cite{Horigane-JPSJ-2007} This increases the number of
magnetic modes in the excitation spectrum and makes it harder to
analyze. We also mention that the magnetic excitations have been
studied in the isostructural half-doped compounds
La$_{3/2}$Sr$_{1/2}$NiO$_4$ (Ref. \onlinecite{Freeman-PRB-2005}) and
La$_{1/2}$Sr$_{3/2}$MnO$_4$ (Ref. \onlinecite{Senff-PRL-2006}).

The paper is organized as follows. After giving details of the
experimental methodology and sample characterization we describe the
results of polarized neutron diffraction measurements designed to
refine the magnetic structure of La$_{3/2}$Sr$_{1/2}$CoO$_4$. We
then present our inelastic neutron scattering measurements which
represent the main body of the work. This is followed by an analysis
of the magnetic spectrum in terms of a generalized spin-wave model.
The paper ends with a discussion of the results and a summary of the
main conclusions from the work.

\section{Experimental Details}

Single crystals of La$_{3/2}$Sr$_{1/2}$CoO$_4$ were grown in Oxford
by the optical floating-zone method in a four-mirror image furnace.
The growth took place in a mixed Ar/O$_2$ atmosphere at a pressure
of 7--9\,bar. The feed and seed rods were scanned at a rate of
3--4\,mm\,hr$^{-1}$ and counter-rotated at 30\,r.p.m. Details of the
preparation and crystal growth procedure are very similar to those
used for La$_{1-x}$Sr$_{x}$CoO$_{3+\delta}$ described in Ref.
\onlinecite{Prabhak-JCG-2005}. Two large crystals were prepared for
the experiments. These were in the form of rods, approximately
10\,mm in diameter and up to 80\,mm in length. Several smaller
crystals prepared under the same conditions were ground up into a
fine powder and used for structural analysis by neutron powder
diffraction. The general materials (GEM) diffractometer at the ISIS
Facility was used to obtain the powder diffraction data. Basic
information on the magnetic response of the crystals was obtained
from magnetization measurements, which were performed on a
superconducting quantum interference device (SQUID) magnetometer
(Quantum Design).

Unpolarized neutron scattering measurements were made on the MAPS
spectrometer at the ISIS Facility. MAPS is a time-of-flight chopper
spectrometer with a large bank of pixelated detectors. The energy of
the incident neutrons is selected by a Fermi chopper, which
transmits pulses of neutrons with an energy bandwidth of typically
3--5\,\%. All the data presented here were obtained with an incident
energy $E_{\rm i}$ of 50\,meV and a chopper frequency of 350\,Hz,
giving an elastic energy resolution of 2\,meV. The crystal used on
MAPS was a single rod with a total mass of 35.5\,g. Of this mass an
estimated 24\,g was in the neutron beam (which was smaller than the
size of the crystal). The crystal was mounted with the rod axis
approximately vertical, and aligned such that the tetragonal $c$
axis was parallel to the incident beam direction. Spectra were
recorded at temperatures of 10\,K, 60\,K and 300\,K. Measurements of
a standard vanadium sample were used to normalize the spectra and
place them on an absolute intensity scale.

Polarized neutron scattering measurements were made on the IN20 and
IN22 triple-axis spectrometers at the Institut Laue-Langevin. On
IN20, two crystals of masses 6.5\,g and 5.5\,g cut from the same rod
were co-aligned with their rod axes parallel. Two settings of the
sample were used, giving access to the $(hk0)$ and $(hhl)$ planes in
reciprocal space [we refer crystallographic notation to the
tetragonal unit cell shown in Fig.\ \ref{fig1}(a)]. For the IN22
experiment we used the 6.5\,g crystal on its own and measured in
just the second orientation. On both IN20 and IN22 we used the (111)
Bragg reflection of Heusler alloy as both monochromator and analyzer
and worked with a fixed final energy $E_{\rm f}$ of 14.7\,meV. A
graphite filter was placed after the sample to suppress higher
harmonics in the scattered beam. Measurements employed uniaxial
polarization analysis,\cite{Moon-PR-1969} and were made with three
configurations of the neutron spin polarization $\bf P$ relative to
the scattering vector $\bf Q$: (a) ${\bf P} \parallel {\bf Q}$, (b)
${\bf P} \perp {\bf Q}$ with $\bf P$ in the scattering plane, and
(c) ${\bf P} \perp {\bf Q}$ with $\bf P$ perpendicular to the
scattering plane. Data were collected in both the spin-flip (SF) and
non-spin-flip (NSF) channels. Details of how the six measurements
$\{({\rm a}), ({\rm b}), ({\rm c})\} \times \{{\rm SF}, {\rm NSF}\}$
can be used to separate the components of a magnetic structure are
given in Ref. \onlinecite{Freeman-PRB-2002}.

\section{Results}

\subsection{Structural characterization}

\begin{figure}
\begin{center}
\includegraphics
[width=8cm,bbllx=0,bblly=0,bburx=416, bbury=244,angle=0,clip=]
{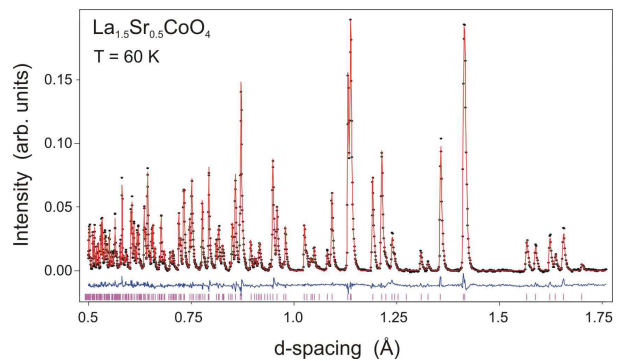} \caption{\label{fig2} (Color online) Part of the
Rietveld refinement for tetragonal
La$_{1.5}$Sr$_{0.5}$CoO$_{4+\delta}$ at 60 K. Circles are data
measured in the backscattering detector bank ($\langle 2\theta
\rangle = 154.5^\circ$) on GEM at ISIS. The solid (red) line shows
the calculated profile fit. Tick marks show the positions of allowed
reflections, and the solid (blue) line below the data shows the
difference between the data and fit. The background fitted in the
refinement was subtracted prior to plotting. The fit parameters are
given in Table \ref{table1}.}
\end{center}
\end{figure}

Neutron powder diffraction data collected on GEM at several
temperatures were analyzed by Rietveld refinement using the GSAS
suite of programs.\cite{GSAS} Data from the three detector banks at
highest scattering-angle were refined simultaneously. Refinements
were made in the space group $I4/mmm$ for the parent structure whose
unit cell is shown in Fig.\ \ref{fig1}(a). Attempts to refine the
structure in the $Fmmm$ space group to allow the possibility of an
orthorhombic distortion did not lead to any
statistically-significant improvements in the quality of the fits.
In addition to the atomic positions shown in Fig.\ \ref{fig1}(a), we
also allowed for the possibility of interstitial oxygen at site
$(0.5,0,0.25)$, where excess oxygen has previously been
found\cite{LeToquin-PhysicaB-2004} in La$_{2}$CoO$_{4+\delta}$. In
order to achieve convergence it was necessary to fix the thermal
parameter ($U_{\rm iso}$) of this third oxygen site to be the same
as that of the second oxygen site.

As an example, Fig.\ \ref{fig2} shows the neutron powder diffraction
pattern and refined profile fit for the data recorded in the
highest-angle detector bank with the sample at a temperature of
60\,K. The parameters for the refinement are listed in Table
\ref{table1} together with those for the other temperatures at which
measurements were made. The data at 2\,K will have contained
magnetic Bragg peaks which were not allowed for in the refinement.
This probably explains the slightly higher value of $R_{\rm wp}$ for
this temperature.  Values for the lattice constants and atomic
positions are in good agreement with previous
data.\cite{Zaliznyak-PRB-2001} To within experimental accuracy the
refinements show no deviation from stoichiometry in the oxygen
content (i.e.\ $\delta =0$), and no evidence for any deviation from
the nominal La:Sr ratio of 3:1.

\begin{table*}[]
\begin{center}
\caption{\label{table1}Structural parameters for
La$_{1.5}$Sr$_{0.5}$CoO$_{4+\delta}$ at temperatures between 2\,K
and 300\,K refined from neutron powder diffraction data. The
refinements were performed in the tetragonal space group $I4/mmm$ of
the parent structure with atomic positions La/Sr 4$e$ $(0,0,z_{\rm
La})$, Co 2$a$ $(0,0,0)$, O(1) 4$c$ $(0.5,0,0)$, O(2) 4$e$
$(0,0,z_{\rm O2})$, O(3) 4$d$ $(0.5,0,0.25)$ in lattice units. The
O(3) position is an interstitial site. $n$ is the occupancy of each
site per formula unit and $U_{\rm iso}$ is the isotropic temperature
factor. The occupancy of the Co site was not refined ($n=1$), and
La/Sr site was constrained such that $n_{\rm La}+n_{\rm Sr}=2$. The
numbers in parentheses are statistical errors in the last digit of
the refined parameters. $R_{\rm wp}$ is the weighted profile
residual function.}
\begin{ruledtabular}
\begin{tabular}{l l l l l l l l}\\[-5pt]
%
\multicolumn{2}{c}{Temperature (K)} & 2 & 60 & 100 & 150 & 200 & 300\\[5pt]
\hline\\
$a=b$ &(\AA)&\footnotesize3.83495(2)&\footnotesize3.83537(2)&\footnotesize3.83665(2)&\footnotesize3.83693(2)&\footnotesize3.83959(2)&\footnotesize3.84080(2)\\
$c$   &(\AA)&\footnotesize12.5235(1)&\footnotesize12.5239(1)&\footnotesize12.5277(1)&\footnotesize12.5287(1)&\footnotesize12.5413(1)&\footnotesize12.5481(1)\\
$V=abc$   &(\AA$^3$)&\footnotesize184.181(2)&\footnotesize184.227(2)&\footnotesize184.406(2)&\footnotesize184.448(2)&\footnotesize184.890(2)&\footnotesize185.107(2)\\[5pt]

La/Sr &$z_{\rm La}$&\footnotesize0.36216(2)&\footnotesize0.36214(2)&\footnotesize0.36216(2)&\footnotesize0.36215(2)&\footnotesize0.36217(3)&\footnotesize0.36216(3)\\
      &$n_{\rm La}$&\footnotesize1.55(7)&\footnotesize1.59(7)&\footnotesize1.52(7)&\footnotesize1.51(7)&\footnotesize1.49(7)&\footnotesize1.47(7)\\
      &$n_{\rm Sr}$&\footnotesize0.45(7)&\footnotesize0.41(7)&\footnotesize0.48(7)&\footnotesize0.49(7)&\footnotesize0.51(7)&\footnotesize0.53(7)\\
      &$U_{\rm iso} \times 100$ (\AA$^2$)&\footnotesize0.219(7)&\footnotesize0.244(7)&\footnotesize0.300(7)&\footnotesize0.307(7)&\footnotesize0.419(8)&\footnotesize0.480(8)\\[5pt]
Co    &$U_{\rm iso} \times 100$ (\AA$^2$)&\footnotesize0.24(3)&\footnotesize0.24(3)&\footnotesize0.30(3)&\footnotesize0.29(3)&\footnotesize0.42(3)&\footnotesize0.43(4)\\[5pt]
O(1)  &$n$&\footnotesize2.01(1)&\footnotesize2.01(1)&\footnotesize2.00(1)&\footnotesize2.00(1)&\footnotesize2.00(1)&\footnotesize1.99(1)\\
      &$U_{\rm iso} \times 100$ (\AA$^2$)&\footnotesize0.50(1)&\footnotesize0.52(1)&\footnotesize0.56(1)&\footnotesize0.57(1)&\footnotesize0.70(1)&\footnotesize0.75(1)\\[5pt]
O(2)  &$z_{O2}$&\footnotesize0.16967(4)&\footnotesize0.16968(4)&\footnotesize0.16968(3)&\footnotesize0.16967(4)&\footnotesize0.16977(4)&\footnotesize0.16981(4)\\
      &$n$&\footnotesize1.98(1)&\footnotesize1.99(1)&\footnotesize1.98(1)&\footnotesize1.98(1)&\footnotesize1.98(1)&\footnotesize1.97(1)\\
      &$U_{\rm iso} \times 100$ (\AA$^2$)&\footnotesize1.02(1)&\footnotesize1.06(1)&\footnotesize1.11(1)&\footnotesize1.12(1)&\footnotesize1.29(1)&\footnotesize1.37(2)\\[5pt]
O(3)  &$n$&\footnotesize0.012(2)&\footnotesize0.008(2)&\footnotesize0.008(2)&\footnotesize0.006(2)&\footnotesize0.004(2)&\footnotesize0.004(2)\\
      &$U_{\rm iso} \times 100$ (\AA$^2$)&\footnotesize1.02(1)&\footnotesize1.06(1)&\footnotesize1.11(1)&\footnotesize1.12(1)&\footnotesize1.29(1)&\footnotesize1.37(2)\\[5pt]
$\delta$& &\footnotesize0.00(2)&\footnotesize0.00(2)&\footnotesize-0.01(2)&\footnotesize-0.02(2)&\footnotesize-0.02(2)&\footnotesize-0.03(2)\\[5pt]
$R_{\rm wp}$&(\%)&\footnotesize3.11&\footnotesize3.04&\footnotesize2.85&\footnotesize2.96&\footnotesize2.88&\footnotesize2.85\\[3pt]
%
\end{tabular}
\end{ruledtabular}
\end{center}
\end{table*}

\subsection{Magnetic structure}

\label{section:magnetic_structure}
\begin{figure}
\begin{center}
\includegraphics
[width=8cm,bbllx=236,bblly=278,bburx=567,bbury=790,angle=0,clip=]
{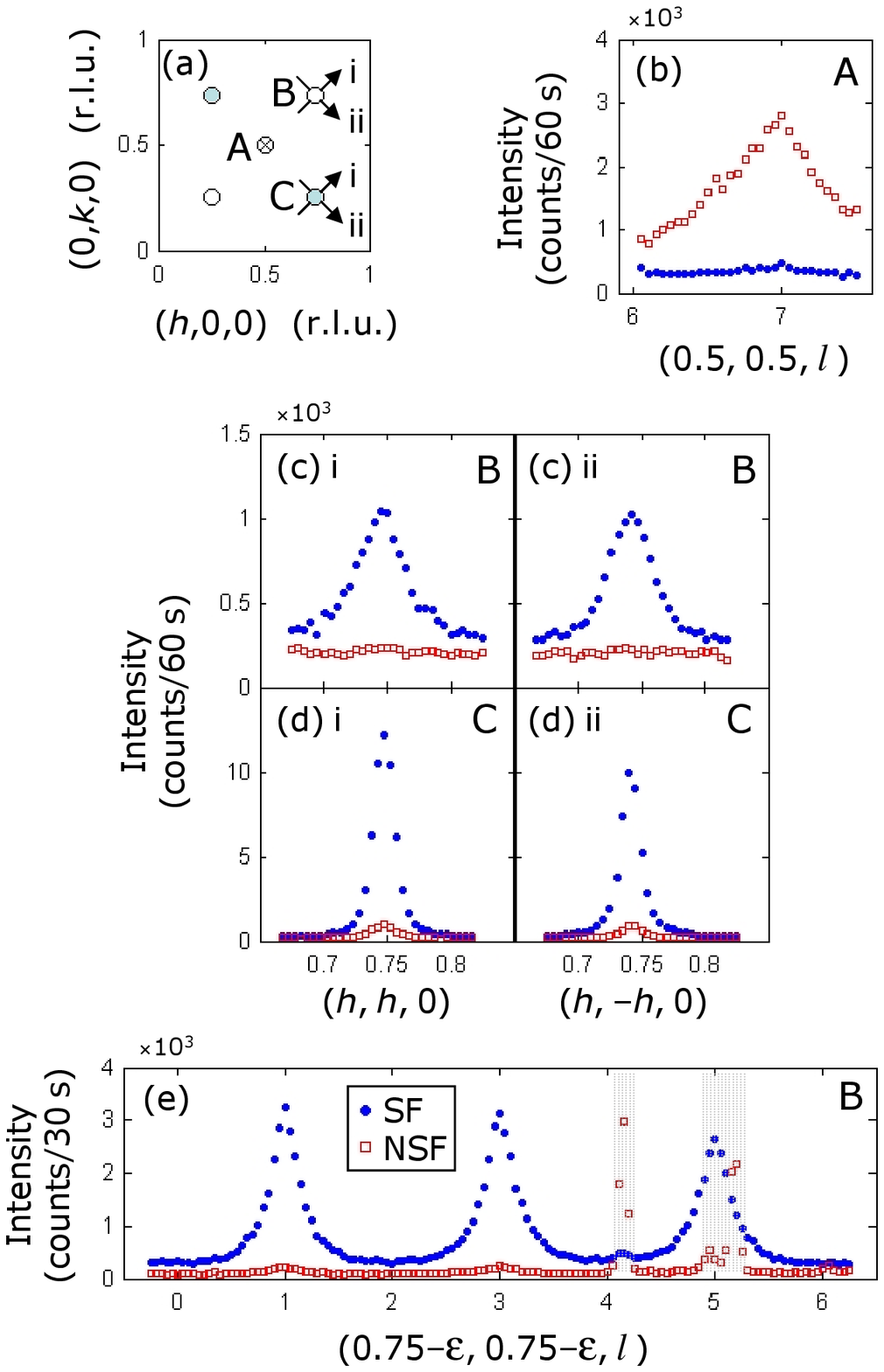} \caption {(Color online) Polarized neutron
diffraction from magnetic and charge orders in
La$_{1.5}$Sr$_{0.5}$CoO$_4$. (a) Diagram of the $(h,k,0)$ reciprocal
lattice plane showing the in-plane wavevectors of the magnetic order
(two twins, unfilled and filled circles marked B and C,
respectively) and charge order (crossed circle marked A). The arrows
indicate scan directions. (b) $l$-dependence of the charge-order
peak at $(0.5,0.5,7)$. (c) and (d) In-plane scans through the
magnetic ordering wavevectors in the directions indicated. (e)
$l$-dependence of the intensity at the magnetic wavevector marked B.
Shaded grey regions contain points contaminated with scattering from
aluminium. All scans shown in this figure were made on IN20 with
${\bf P}
\parallel {\bf Q}$ and recorded in both the spin-flip (SF, filled
blue circles) and non-spin-flip (NSF, open red squares) channels. No
corrections have been applied for the imperfect neutron
polarization. The sample temperature was 2\,K.} \label{fig3}
\end{center}
\end{figure}

We first review the main
characteristics of the magnetic and charge order in
La$_{1.5}$Sr$_{0.5}$CoO$_{4}$ with the aid of the polarized-neutron
diffraction measurements presented in Fig.\ \ref{fig3}. All the
scans shown in this figure contain raw data recorded at 2\,K on IN20
with the neutron polarization parallel to the scattering vector
(${\bf P}
\parallel {\bf Q}$). In this configuration the spin-flip
scattering is entirely magnetic and the non-spin-flip scattering is
entirely non-magnetic (i.e.\ structural). Since the neutron
polarization is not perfect (the flipping ratio was $\sim 20$) there
is a small amount of leakage from one channel to the other.

Figure~\ref{fig3}(a) is a diagram of part of the $(h,k,0)$ plane in
reciprocal space showing the in-plane wavevectors corresponding to
the magnetic and charge order. Representative scans along various
in-plane and out-of-plane directions are shown in Figs.\
\ref{fig3}(b)--(e). These data confirm that the charge order
scattering is peaked at wavevectors ${\bf Q}_{\rm co}=(h + 0.5,k+
0.5,l)$, with $h$, $k$ and $l$ integers, and that the magnetic order
is characterized by four slightly-incommensurate wavevectors ${\bf
Q}_{\rm m}=(h + 0.5,k+ 0.5,l) \pm (0.25-\epsilon ,0.25-\epsilon,0)$
and $(h + 0.5,k+ 0.5,l) \pm (-0.25+\epsilon ,0.25-\epsilon,0)$,
where $h$ and $k$ are integers and $l$ is an odd integer for the
first pair of ${\bf Q}_{\rm m}$ and an even integer for the second
pair. In our crystal $\epsilon=0.005\pm 0.001$. The peaks are
considerably sharper in scans parallel to the $ab$ plane [e.g.\
Figs.\ \ref{fig3}(c) and (d)] than in scans made in the out-of-plane
direction [Figs.\ \ref{fig3}(b) and (e)], but in all directions the
peak widths are broader than the resolution. The correlation lengths
for the charge and magnetic orders estimated from $\xi = 1/\Gamma$,
where $\Gamma$ is the half-width at half-maximum, are $\xi_{\rm
co}^{ab} = 23\pm 2$\,{\AA}, $\xi_{\rm co}^{c} = 8\pm 1$\,{\AA},
$\xi_{\rm m}^{ab} = 52 \pm 2$\,{\AA} and $\xi_{\rm m}^{c} = 12 \pm
1$\,{\AA}. These values are consistent with those reported by
Zaliznyak {\it et al.}\cite{Zaliznyak-PRL-2000} apart from $\xi_{\rm
m}^{ab}$, which we find to be $\sim 30$\% smaller. This could be
because we did not attempt to correct the peak width for
experimental resolution.

The two in-plane scans shown in Fig.~\ref{fig3}(c) are at $l=0$,
which is a minimum of the intensity modulation along $(0,0,l)$ ---
see Fig.~\ref{fig3}(e). The widths of the peaks in
Fig.~\ref{fig3}(c) are roughly twice those in Fig.~\ref{fig3}(d)
which are at a maximum of the $(0,0,l)$ intensity modulation. This
suggests that in addition to the dominant magnetic order there also
exists a small volume fraction of magnetically ordered regions with
an average size of $\sim 25$\,{\AA} in the $ab$ plane which are not
correlated to the magnetic order on the adjacent layers.

It is reasonable to assume that the short correlation lengths in the
$c$ direction are caused by the existence of different stacking
sequences with similar energy. Since both the charge and magnetic
diffraction peaks are found at integer $l$ the majority stacking is
periodic in the lattice. However, there are several ways in which
the structure on adjacent layers can be related. To distinguish
these experimentally let us consider the case where the magnetic
structure is collinear, so that the spin--charge order on one layer
(say $z=0$) can be related to that on the adjacent layer
($z=\frac{1}{2}$) by a translation ${\bf t}$. The structure factor
for the spin and charge diffraction peaks then contains a factor
$1+\exp(i{\bf Q}\cdot {\bf t}) = 1 + \exp\{2\pi i
(ht_x+kt_y+lt_z)\}$, where $t_x, t_y, t_z$ are the components of
$\bf t$ written as fractional coordinates along the crystallographic
$a$, $b$ and $c$ axes.

We consider first the case ${\bf t} = (-0.5,0.5,0.5)$. For the
structure shown in Fig.\ \ref{fig1}(b) this stacking leads to
systematic absences at magnetic ordering wavevectors of the type
$(0.25,0.25,l)$, $(0.75,0.75,l)$, etc, when $l$ is odd and for
magnetic wavevectors of the type $(0.25,0.75,l)$, $(0.75,0.25,l)$,
etc [associated with the twin obtained by rotating the structure in
Fig.\ \ref{fig1}(b) by 90\,deg
--- see Fig.\ \ref{fig3}(a)], when $l$ is even. These predictions are inconsistent with experiment
--- see Fig.\ \ref{fig3}(e). We next consider ${\bf t} = (0.5,0.5,0.5)$.
In this case there are no magnetic absences, again inconsistent with
experiment. We finally consider ${\bf t} = (1.5,0.5,0.5)$. For this
case the systematic absences are at $(0.25,0.25,l)$, etc, when $l$
is even, and at $(0.25,0.75,l)$, etc, when $l$ is odd. This is
consistent with the observations. The twinning ensures that there
are no absences in the charge-order peaks, again consistent with
experiment.\cite{Zaliznyak-PRL-2000} We conclude, therefore, that
the most likely stacking vector is ${\bf t} = (1.5,0.5,0.5)$ (or its
equivalent). This vector is shown in Fig.\ \ref{fig1}(b). With this
stacking, spins in the $z=\frac{1}{2}$ layer are antiparallel to the
closest spins in the $z=0$ and $z=1$ layers.

The fact that the magnetic and charge correlation lengths along the
$c$ axis are so short suggests that the majority stacking is only
slightly more favorable energetically than other possible stacking
sequences. Therefore, the propagation of the structure along the $c$
axis is presumably interrupted frequently by stacking faults in
which adjacent layers are related by different ${\bf t}$ vectors.

\begin{figure}
\begin{center}
\includegraphics
[width=8cm,bbllx=37,bblly=305,bburx=380,bbury=535,angle=0,clip=]
{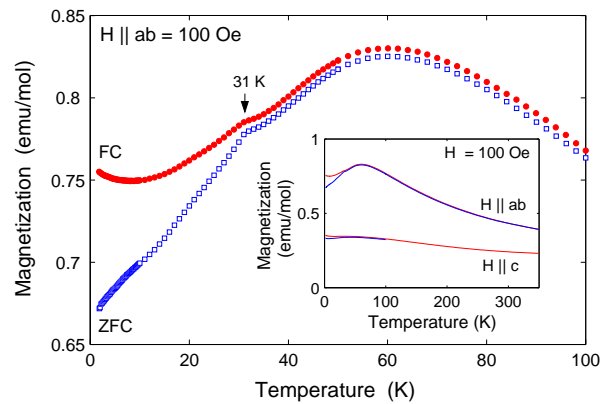} \caption {(Color online) Temperature dependence
of the magnetization of La$_{1.5}$Sr$_{0.5}$CoO$_4$. The main frame
shows measurements made with the applied field $H$ of strength
100\,Oe ($\mu_{0}H = 0.01$\,T) applied parallel to the $ab$ plane.
Red filled circles and blue open squares show data from field-cooled
(FC) and zero-field-cooled (ZFC) measurements, respectively. Inset:
data up to $T=350$\,K showing measurements with both $H\parallel ab$
and $H\parallel c$.} \label{fig4}
\end{center}
\end{figure}

Let us now turn to the temperature evolution of the magnetic order.
In Fig.\ \ref{fig4} we show the magnetization of
La$_{1.5}$Sr$_{0.5}$CoO$_{4}$ as a function of temperature, recorded
with a relatively low measuring field $H$ of 100\,Oe. The data are
in very good quantitative agreement with the measurements reported
in Refs.\ \onlinecite{Moritomo-PRB-1997} and
\onlinecite{Hollmann-NJP-2008}, both of which employed a
significantly higher measuring field. The magnetization exhibits
strong $XY$-like anisotropy, as shown in the inset of Fig.\
\ref{fig4}, and there is a broad maximum in the in-plane response
centered at about 60\,K. At lower temperatures there is a splitting
between the field-cooled (FC) and zero-field-cooled (ZFC) data. The
splitting is largest for $H \parallel ab$.

Our measurements, however, reveal an additional feature. At
approximately 31\,K there is a sharp kink in the data. This kink
marks the temperature below which the FC--ZFC splitting begins to
open up most rapidly. The observation of this kink prompted us to
perform polarized-neutron diffraction measurements to investigate
whether there might be a change in the magnetic structure at 31\,K.

We followed the approach described in Ref.\
\onlinecite{Freeman-PRB-2002} (see also Ref.
\onlinecite{Helme-DPhil-2006}). SF and NSF intensities were recorded
at two magnetic Bragg peaks using three orthogonal directions of the
neutron polarization $\bf P$. Measurements were made on both IN20
and IN22, and corrections were applied to the measured intensities
to compensate for the imperfect polarization. The two Bragg peaks
used were ${\bf Q}_1 = (0.25+\epsilon,0.25+\epsilon,7)$ and ${\bf
Q}_2 = (1.25+\epsilon,1.25+\epsilon,1)$. These were chosen because
they make small angles (less than 10\,deg) to the $c$ axis and
$[110]$ direction, respectively, which reduces the uncertainty in
the values of the spin components derived by this method. For
example, if ${\bf Q}_1$ could be chosen exactly parallel to $c$ then
the Bragg peak intensities measured at ${\bf Q}_1$ would be
independent of the magnetic component along the $c$ axis.

Because ${\bf Q}_1$ and ${\bf Q}_2$ lie in the $(h,h,l)$ plane the
magnetic scattering at these wavevectors is naturally described in
terms of the intensities scattered by the projection of the ordered
moments along the orthogonal directions $[1,1,0]$, $[1,-1,0]$ and
$[0,0,1]$. We call these intensities $I_{110}$, $I_{1\bar{1}0}$ and
$I_{c}$, respectively. The expressions in Table I of Ref.\
\onlinecite{Freeman-PRB-2002} can be used to determine the ratios
$I_{110}/I_{1\bar{1}0}$ and $I_{c}/(I_{110}+I_{1\bar{1}0})$ from the
sets of measurements at ${\bf Q}_1$ and ${\bf Q}_2$. The latter
ratio was found to be less than 0.01 at all temperatures. The
assumption that the intensities are proportional to the squares of
the ordered moments constrains the angle of the moments to the $ab$
plane to $<5$\,deg. Given the strong planar anisotropy it is safe to
assume that the moments lie in the $ab$ plane.


\begin{figure}
\begin{center}
\includegraphics
[width=8cm,bbllx=152,bblly=268,bburx=445,bbury=515,angle=0,clip=]
{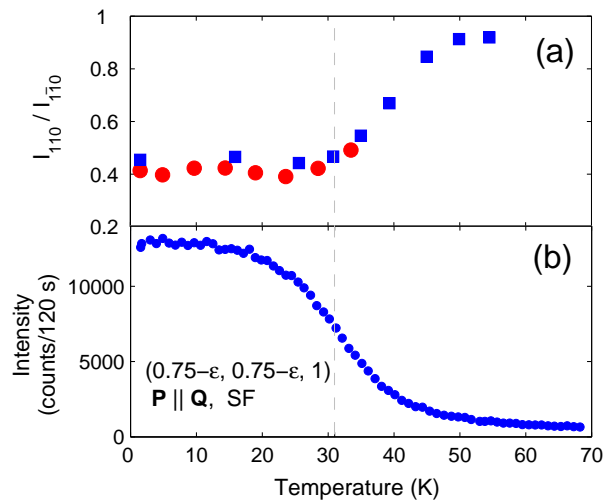} \caption {(Color online) Development of magnetic
order in La$_{1.5}$Sr$_{0.5}$CoO$_{4}$. (a) Ratio of the intensities
scattered by the projection of the ordered moments along the $[110]$
and $[1\bar{1}0]$ directions. Data are from IN20 (filled red
circles) and IN22 (filled blue squares). (b) Intensity of the
$(0.75-\epsilon,0.75-\epsilon,1)$ magnetic Bragg peak. The broken
grey line marks the position of the kink in the magnetization (see
Fig.\ \ref{fig4}).} \label{fig5}
\end{center}
\end{figure}

In Fig.\ \ref{fig5}(a) we plot the ratio $I_{110}/I_{1\bar{1}0}$ as
a function of temperature. At temperatures below 30\,K,
$I_{110}/I_{1\bar{1}0}$ is approximately constant with a value of
0.4. On warming above 30\,K this ratio gradually increases until at
$\sim 50$\,K it approaches 1. This indicates one of two things.
Either (i) there are two (or more) spin domains which contribute to
each magnetic Bragg peak, and on cooling one of these domains
becomes preferentially populated, or (ii) there is a spin-canting
transition reminiscent of that proposed to occur in the
isostructural layered
nickelates.\cite{Freeman-PRB-2002,Lee-PRB-2001,Freeman-PRB-2004,Giblin-PRB-2008}
We consider these possibilities further in Section \ref{Discussion}

Figure~\ref{fig5}(b) shows the temperature dependence of the
intensity of the $(0.75-\epsilon,0.75-\epsilon,1)$ magnetic peak.
The slow increase in intensity on cooling starts in the vicinity of
the broad hump in the magnetization at $\sim 60$\,K, confirming that
the hump is associated with the build-up of magnetic correlations.
We have indicated on Figs.\ \ref{fig5}(a) and (b) the temperature at
which the kink is observed in the magnetization. There is no obvious
anomaly in the magnetic peak intensity at this temperature, but the
data in Fig.\ \ref{fig5}(a) indicate that the 31\,K kink marks the
temperature below which the magnetic structure stops changing.

\subsection{Magnetic excitations}
\label{section:magnetic_excitations}

\begin{figure}
\begin{center}
\includegraphics
[width=8cm,bbllx=61,bblly=367,bburx=487,bbury=771,angle=0,clip=]
{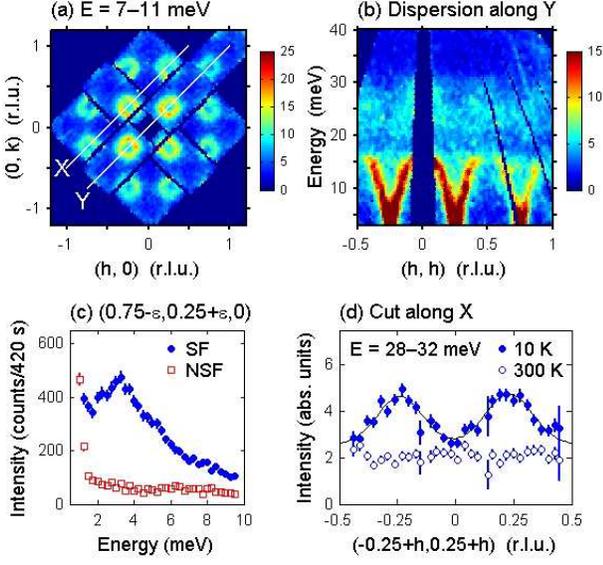} \caption {(Color online) Spin excitation spectrum
of La$_{1.5}$Sr$_{0.5}$CoO$_{4}$. (a) Map of the intensity within
the $(h,k)$ plane averaged over the energy range $7-11$\,meV,
measured with a sample temperature of $T=10$\,K . The circles
centered on the magnetic ordering wavevectors are due to scattering
from dispersive magnetic excitations. (b) Energy--$\bf Q$ slice
showing the dispersion along the line Y in (a). (c) Energy scan at
the magnetic ordering wavevector $(0.75-\epsilon,0.25+\epsilon,0)$
with neutron polarization analysis to separate the magnetic
scattering in the spin-flip (SF) channel from the non-magnetic
scattering in the non-spin-flip (NSF) channel. The data were
obtained on IN20 at $T=2$\,K and reveal an energy gap of $\sim
3$\,meV in the magnetic spectrum. (d) Scans along the line X in (a)
at temperatures of 10\,K and 300\,K averaged over the energy range
$28-32$\,meV confirming the existence of magnetic modes at this
energy. The data in (a), (b) and (d) were recorded on MAPS, and the
intensity is in units of mb~sr$^{-1}$~meV$^{-1}$~f.u.$^{-1}$, where
``f.u." stands for ``formula unit" (of
La$_{1.5}$Sr$_{0.5}$CoO$_{4}$).} \label{fig6}
\end{center}
\end{figure}

\begin{figure*}
\begin{center}
\includegraphics
[width=16cm,bbllx=61,bblly=381,bburx=589,bbury=750,angle=0,clip=]
{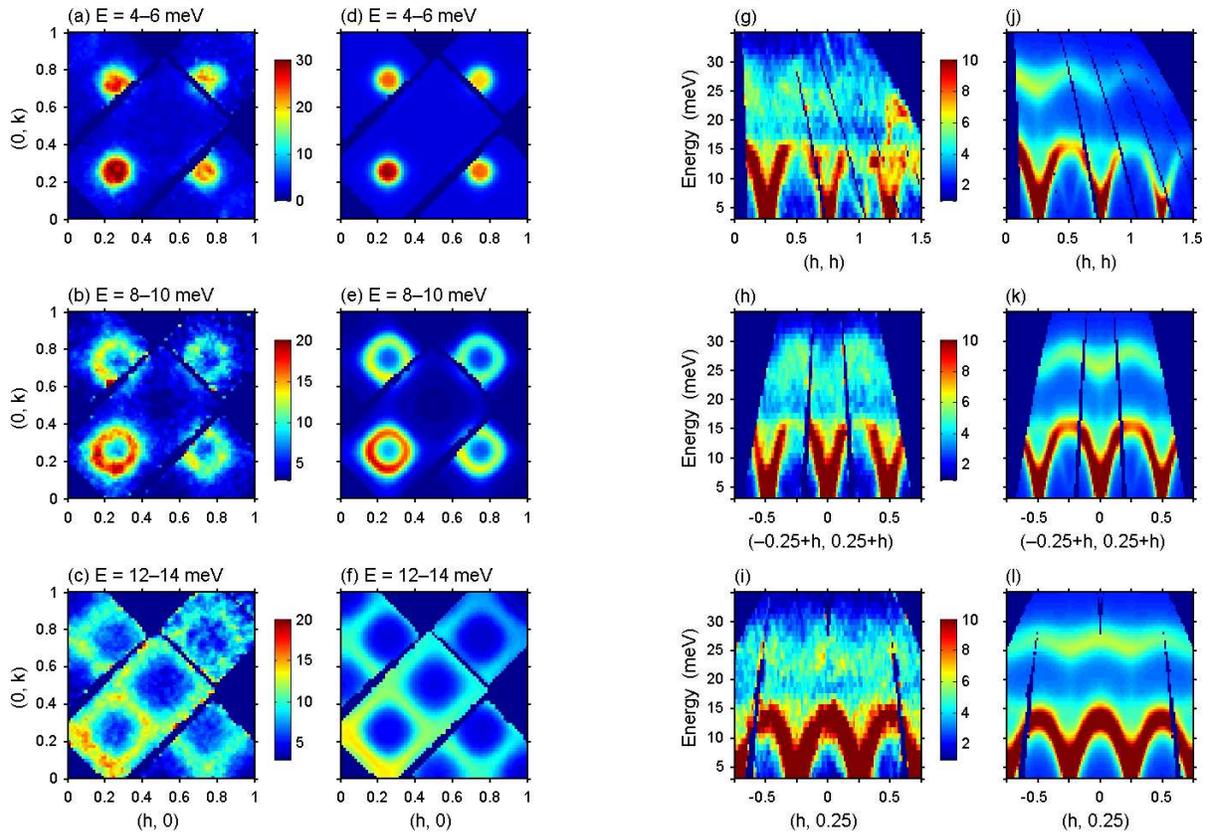} \caption {(Color online) Intensity maps of
measured and simulated neutron scattering from
La$_{1.5}$Sr$_{0.5}$CoO$_{4}$. The data are from MAPS, and the
simulations are obtained from the many-level spin-wave model
discussed in the text with Co$^{2+}-$Co$^{2+}$ exchange interactions
$J=1.4$\,meV and $J_1=J_2=0$ [see Fig.\ \ref{fig1}(b)]. (a)--(f)
Constant-energy slices averaged over the energy ranges indicated.
(g)--(l) Energy--$\bf Q$ slices showing the dispersion along the
three symmetry directions indicated. The intensity scale is in units
of mb\,sr$^{-1}$\,meV$^{-1}$\,f.u.$^{-1}$.} \label{fig7}
\end{center}
\end{figure*}

Figure \ref{fig6} provides an overview of the magnetic spectrum. The
intensity map presented in Fig.\ \ref{fig6}(a) is a slice from a
MAPS data volume, averaged over the energy range $7-11$\,meV and
projected onto the $(h,k)$ plane. The image shows dispersive
magnetic excitations emerging from the magnetic ordering
wavevectors. The rings of scattering correspond to the intersection
of the constant-energy slice plane with the spin-wave cones. Figure
\ref{fig6}(b) is an energy-wavevector slice through the same data
volume showing the dispersion along a line parallel to $(h,h,0)$.
The spectrum is dominated by an intense band of scattering extending
up to 16\,meV. There is also a weaker band of scattering in the
range $20-30$\,meV which is more diffuse than that of the lower band
but which disperses with the same period.

Figure \ref{fig6}(c) is an energy scan performed on IN20 with the
wavevector fixed at the magnetic ordering wavevector ${\bf Q}_m =
(0.75-\epsilon,0.25+\epsilon,0)$. The two sets of points are the
neutron spin-flip (SF) and non-spin-flip (NSF) signals. In the ${\bf
P} \parallel {\bf Q}$ configuration used for this measurement the SF
channel contains only magnetic scattering, so the large signal in
the SF data is scattering from magnetic excitations and reveals a
gap of approximately 3\,meV in the low-energy spin-wave band. Having
observed the gap we performed constant-energy scans (not shown)
along $(0.75-\epsilon,0.75-\epsilon,l)$ from $l=0$ to 2 at energies
of 2\,meV and 4\,meV, i.e.\ just below and just above the gap. At
2\,meV there remained a modulation in the magnetic scattering along
the scan with a broad maximum at the magnetic Bragg peak position
$(0.75-\epsilon,0.75-\epsilon,1)$, whereas at 4\,meV the scan was
featureless to within the experimental precision of $\sim 5$\%. This
implies that for energies above 4\,meV the magnetic dynamics are
completely uncorrelated along the $c$ axis and the spectrum can be
considered as two-dimensional.

To help understand the origin of the gap we performed neutron
polarization analysis at energies of 2\,meV and 4\,meV. For each
energy we recorded the signal in the same three polarization
channels and at the same two wavevectors ${\bf Q}_1 =
(0.25+\epsilon,0.25+\epsilon,7)$ and ${\bf Q}_2 =
(1.25+\epsilon,1.25+\epsilon,1)$ as used to analyze the magnetic
order (see Section \ref{section:magnetic_structure}). The signal in
each polarization channel can be written in terms of the response
functions $S^{110}({\bf Q}_j,\omega)$, $S^{1\bar{1}0}({\bf
Q}_j,\omega)$ and $S^{zz}({\bf Q}_j,\omega)$ for magnetic
fluctuation components parallel to $[110]$, $[1\bar{1}0]$, and
parallel to the $c$ axis ($z$ direction), respectively. Here,
$\omega$ is the neutron energy transfer and $j=1,2$ indexes the two
magnetic wavevectors. Applying the same analysis as used to separate
diffraction from different components of the ordered
moments\cite{Freeman-PRB-2002,Helme-DPhil-2006} we find at $T =
2$\,K,
\begin{eqnarray}
S^{zz}({\bf Q}_j,2{\rm \,meV})/S^{1\bar{1}0}({\bf Q}_j,2{\rm \,meV})
& =
& -0.04 \pm 0.06 \nonumber \\
S^{zz}({\bf Q}_j,4{\rm \,meV})/S^{1\bar{1}0}({\bf Q}_j,4{\rm \,meV})
& = & -0.02 \pm 0.04 \nonumber \\
S^{110}({\bf Q}_j,2{\rm \,meV})/S^{1\bar{1}0}({\bf Q}_j,2{\rm
\,meV}) &
= & 1.19 \pm 0.07 \nonumber \\
S^{110}({\bf Q}_j,4{\rm \,meV})/S^{1\bar{1}0}({\bf Q}_j,4{\rm
\,meV}) & = & 1.00 \pm 0.05\ . \nonumber
\end{eqnarray}
The first two ratios show that the spin fluctuations at both 2\,meV
and 4\,meV are restricted to the $ab$ plane. This is consistent with
the strong $XY$-like anisotropy of this system and implies that the
gap is due to a small single-ion or exchange anisotropy within the
$ab$ plane. The third and fourth ratios show that the strengths of
the fluctuations in the $[110]$ and $[1\bar{1}0]$ directions are
roughly equal in this energy range.

In Fig.\ \ref{fig6}(d) we show cuts through the MAPS data averaged
over the energy range 28\,meV to 32\,meV, which is near the top of
the weak upper band of excitations. The scan at 10\,K shows a
sinusoidal intensity modulation along the cut direction [line X in
Fig.\ \ref{fig6}(a)], whereas the scan at 300\,K is flat and always
below the 10\,K data. The observation that the signal goes away on
raising the temperature from 10\,K to 300\,K shows conclusively that
it is magnetic in origin, because the scattering from phonons would
increase in intensity.

To give a more comprehensive picture of the magnetic excitation
spectrum we present in Fig.\ \ref{fig7} a further series of slices
through the MAPS data. The data have been averaged over
symmetry-equivalent directions to improve the statistics, and are
plotted with respect to the two-dimensional reciprocal lattice of
the CoO$_2$ layers indexed by $(h,k)$. In the time-of-flight method
the out-of-plane wavevector component $l$ varies with energy and
also varies across the detector for a fixed energy. However, because
the magnetic dynamics is two-dimensional (for energies greater than
$\sim$4\,meV) the magnetic dispersion does not depend on $l$ and the
intensity of the spectrum varies only slowly with $l$. The $l$
dependence of the intensities of the modes is correctly taken into
account in the models presented in Section~\ref{analysis}.

Figures \ref{fig7}(a)--(c) are constant-energy slices averaged over
a 2\,meV interval centered on 5\,meV, 9\,meV and 13\,meV. These
illustrate how the lower spin-wave band disperses away from the
magnetic zone centers. Figures \ref{fig7}(g)--(i) display three
energy--wavevector slices taken along different symmetry directions
in the Brillouin zone. The intense lower band and weak upper band
are clearly seen in each of these slices.

\begin{figure}
\begin{center}
\includegraphics
[width=8cm,bbllx=17,bblly=19,bburx=435,bbury=757,angle=0,clip=]
{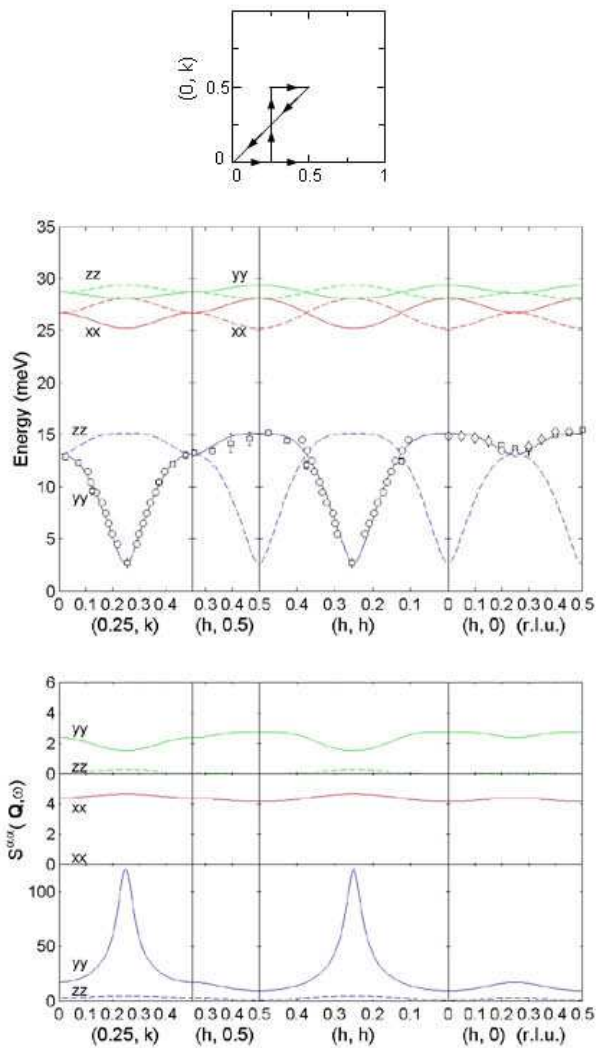} \caption {(Color online) Dispersion of magnetic
excitations in La$_{1.5}$Sr$_{0.5}$CoO$_{4}$. The diagram at the top
shows the path in reciprocal space along which the dispersion is
plotted. The middle figure shows the measured and calculated
dispersion. Symbols are from fits to the experimental data. Circles
are from constant-energy cuts, squares and diamonds are from
constant-wavevector cuts. The lines are calculated from the
many-level spin--orbital model with Co$^{2+}-$Co$^{2+}$ exchange
interactions $J=1.4$\,meV and $J_1=J_2=0$ [see Fig.\ \ref{fig1}(b)].
The lower figure displays the calculated response functions
$S^{\alpha \alpha}({\bf Q},\omega)$ for each mode shown in the
dispersion plot. The labels on the curves give the $\alpha \alpha$
components for the case where the ordered moments point along $x$.}
\label{fig8}
\end{center}
\end{figure}

To facilitate comparison with models we took cuts through the MAPS
data set along several symmetry directions and extracted points
describing the dispersion. The magnetic dispersion obtained from the
MAPS data is displayed in the upper panel of Fig.\ \ref{fig8}. The
points are obtained either from wavevector cuts at fixed energy, or
from energy cuts at fixed wavevector. Peaks in the cuts were fitted
with Lorentzian functions on a linear background. The points
representing the energy gap at the magnetic zone centres were
estimated from the data in Fig.\ \ref{fig6}(c).

\section{Analysis}
\label{analysis}

\begin{figure}
\begin{center}
\includegraphics
[width=8cm,bbllx=137,bblly=128,bburx=557,bbury=390,angle=0,clip=]
{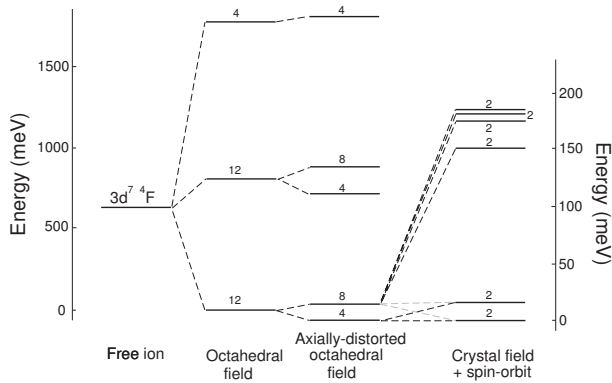} \caption {Single-ion energy levels of the $^4F$
term of Co$^{2+}$ in La$_{1.5}$Sr$_{0.5}$CoO$_{4}$, showing how the
term splits when successively smaller interactions are added. The
splitting caused by the axially-distorted crystal field is obtained
with coefficients $B_2^0 = 13$\,meV, $B_4^0 = -1.4$\,meV and $B_4^4
= -8$\,meV for the Stevens operator equivalents $O_2^0, O_4^0$ and
$O_4^4$. These values are approximately twice those calculated from
the point-charge model. The right-most scheme includes spin--orbit
coupling and shows the splitting of the lowest orbital quasi-triplet
on an expanded vertical scale. The spin--orbit coupling parameter is
$\lambda = -18.7$\,meV. } \label{fig9}
\end{center}
\end{figure}
The essential physics governing the magnetic properties of
La$_{1.5}$Sr$_{0.5}$CoO$_{4}$ can be understood with reference to
Fig.\ \ref{fig9}, which is an energy-level diagram for the high-spin
state of Co$^{2+}$ (3$d^7$, $L=3$, $S=3/2$) in an axially-distorted
octahedral crystal field. A perfect octahedral field splits the
orbital levels into two triplets and a singlet, and the small axial
elongation of the octahedron splits each triplet into a singlet and
a higher-lying doublet. Inclusion of spin--orbit coupling lifts the
fourfold spin degeneracy of each of the orbital levels, splitting
the ground state into two doublets. For realistic crystal field and
spin--orbit interactions the two doublets are separated by
approximately 25\,meV, as we show later. Because the orbital ground
state in the main octahedral field is a quasi-triplet it contains
significant unquenched orbital angular momentum which is responsible
for the strong planar anisotropy. As we shall see, the magnetic
spectrum in the energy range probed in this study involves the
lowest two doublets and is strongly influenced by the in-plane
anisotropy and the unquenched orbital angular momentum.

The neutron scattering cross section for spin-only scattering, or
for spin and orbital scattering in the dipole approximation, may be
written \cite{Squires}
\begin{eqnarray}
\frac{{\rm d}^2\sigma}{{\rm d}\Omega{\rm d}E_{\rm f}} & = &
\frac{k_{\rm f}}{k_{\rm i}}\left(\frac{\gamma r_0}{2}
\right)^{\hspace{-2pt}2} f^2(Q)\exp(-2W) \nonumber\\[5pt] & & \hspace{5pt}\times \sum_{\alpha
\beta}(\delta_{\alpha
\beta}-\hat{Q}_{\alpha}\hat{Q}_{\beta})S^{\alpha\beta}({\bf
Q},\omega),\label{Eq1}
\end{eqnarray}
where $(\gamma r_0/2)^2 = 72.8$\,mb, $f(Q)$ is the magnetic form
factor of Co$^{2+}$, $\exp(-2W)$ is the Debye--Waller factor which
is close to unity at low temperatures, $\hat{Q}_{\alpha}$ is the
$\alpha$ component of a unit vector in the direction of $\bf Q$, and
$S^{\alpha\beta}({\bf Q},\omega)$ is the response function
describing $\alpha\beta$ magnetic correlations. In calculating the
cross-section we average over an assumed 50:50 ratio of the two
magnetic domains related to one another by a 90\,deg rotation.

Putting the orbital angular momentum to one side for a moment, we
can as a first approximation follow the standard route and attempt
to describe the low-energy magnetic dynamics in the
antiferromagnetic phase in terms of linear spin-wave excitations of
an effective spin--$\frac{1}{2}$ model for the ground state doublet.
We assume the collinear magnetic structure shown in Fig.~\ref{fig1}
with spins along the $a$ axis ($\phi=0$). We consider a spin-only
Hamiltonian and incorporate the magnetic anisotropy via anisotropic
exchange interactions between effective $S=\frac{1}{2}$ spins. The
Hamiltonian may be written
\begin{equation}
{\mathcal H} = \sum_{\langle jk\rangle}
\sum_{\alpha}J_{jk}^{\alpha}\,S_j^{\alpha} S_k^{\alpha}.\label{Eq2}
\end{equation}
The first summation is over Co$^{2+}-$Co$^{2+}$ pairs with each pair
counted only once, and the second summation is over the spin
components $\alpha = x,y,z$.  The $J_{jk}^{\alpha}$ are the exchange
parameters. We include only $J$, $J_1$ and $J_2$ as defined in Fig.\
\ref{fig1}. $J$ acts in a straight line through the Co$^{3+}$ site,
whereas $J_1$ and J$_2$ have 90\,deg paths. As we shall see, $J_1$
and $J_2$ are very much smaller than $J$, and so as a simplification
we take $J_1$ and $J_2$ to be isotropic. For later convenience we
write the anisotropy in $J$ in the form $J^x = J(1+\varepsilon)$,
$J^y = J$ and $J^z = J(1-\delta)$ so that $\varepsilon$ and $\delta$
parameterize the degree of in-plane and out-of-plane anisotropy,
respectively.

The theoretical expressions for the spin-wave dispersion and
response functions are given in the Appendix,
Eqs.~(\ref{App-eq2})--(\ref{App-eq5}). For each wavevector there are
two modes which are degenerate in the absence of anisotropy.
Inclusion of easy-plane anisotropy lifts the degeneracy so that one
mode ($\omega_1$) is associated with in-plane fluctuations and the
other mode ($\omega_2$) is associated with out-of-plane
fluctuations. At the magnetic Bragg peak wavevectors $(0.25,0.25)$,
$(0.75,0.75)$, etc. the out-of-plane mode is gapped while the
in-plane mode is gapless. The addition of a small in-plane
anisotropy gaps the in-plane mode too. In our case we require a
large out-of-plane anisotropy to describe the data. To see this we
refer to the scattering intensity maps, e.g. Fig.~\ref{fig7}(g), and
the dispersion data in Fig.~\ref{fig8}. If the  anisotropy were
small then the dispersion would reach a maximum at, or very close
to, the magnetic zone boundaries $(0.125,0.125)$, $(0.375,0.375)$,
etc. However, the measured dispersion reaches a maximum not at these
positions but at the zone centres $(0,0)$, $(0.5,0.5)$, etc. This
behavior can be reproduced in the linear spin-wave model only if
there is substantial easy-plane anisotropy, i.e. if $J^z$ is
significantly less than $J^x$ and $J^y$.

We found that the observed dispersion for the lower energy modes can
be described very well throughout the Brillouin zone by the
effective spin--$\frac{1}{2}$ spin-wave model with parameters
$SJ=3.3$\,meV, $J_1=J_2=0$, $\delta=0.65$ and $\varepsilon=0.03$.
The zero values of the 90\,deg exchange parameters $J_1$ and $J_2$
means that the magnetic structure consists of two uncoupled,
interpenetrating, square-lattice antiferromagnets with spacing $2a$
on the CoO$_2$ layers. This is surprising and has important
implications for the stability of the magnetic structure, so let us
examine the evidence for this finding carefully.


\begin{figure}
\begin{center}
\includegraphics
[width=7.5cm,bbllx=140,bblly=298,bburx=391,bbury=487,angle=0,clip=]
{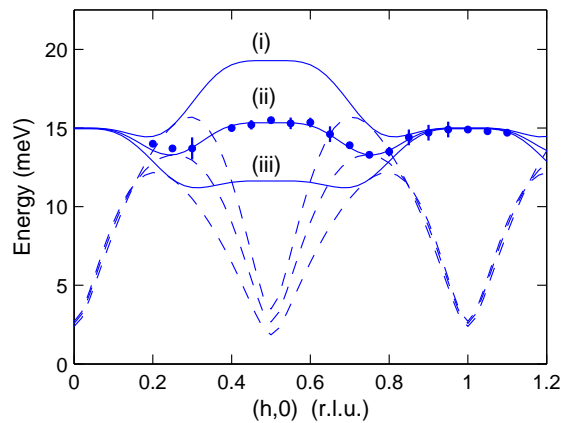} \caption {(Color online) Magnetic dispersion
parallel to the $(h,0)$ direction in La$_{1.5}$Sr$_{0.5}$CoO$_{4}$.
The symbols are points determined from energy cuts at constant
wavevector through the MAPS data. Full and broken lines are in-plane
and out-of-plane modes of the effective spin--$\frac{1}{2}$
ground-state doublet calculated from Eq.~(\ref{Eq2}) by linear
spin-wave theory with the following exchange parameters: (i)
$SJ=3.68$\,meV, $J_1=J_2=-J/2$; (ii) $SJ=3.23$\,meV,
$SJ_1=SJ_2=-0.15$\,meV; (iii) $SJ=2.85$\,meV, $J_1=J_2=J/2$. The
parameters in (ii) have been adjusted to give the best fit to the
data. In each case the exchange-anisotropy parameters were
$\varepsilon = 0.03$ and $\delta = 0.65$.} \label{fig10}
\end{center}
\end{figure}

First of all we show in Fig.~\ref{fig10} the measured dispersion
along the $(h,0)$ line in reciprocal space. If the tetragonal
symmetry is preserved in the charge-ordered phase then one expects
$J_1 = J_2$, which will tend to frustrate the magnetic order. If
$J_1$ and $J_2$ are both antiferromagnetic then the effect on the
dispersion relation is to soften the in-plane mode at the zone
boundaries along the $(h,0)$ and $(0,k)$ directions. If $J_1$ and
$J_2$ are both ferromagnetic then the softening is instead at the
reciprocal lattice vectors. In both cases, therefore, we should
observe a difference between the magnon energy at $(0.5,0)$ and at
$(1,0)$. It can be seen from the data points in Fig.~\ref{fig10}
that if such a difference exists then it is very small. To be more
quantitative we plot on Fig~\ref{fig10} the dispersion curves
calculated from linear spin-wave theory for the cases $J_1=J_2=J/2$
($J$, $J_1$, $J_2$ all antiferromagnetic) and $J_1=J_2=-J/2$ ($J$
antiferromagnetic, $J_1$, $J_2$ ferromagnetic), adjusting the value
of $J$ to accord with the data at $(1,0)$. These curves fall
respectively well below and well above the experimental data,
showing that the magnitudes of $J_1$ and $J_2$ must be considerably
less than half that of $J$. The third curve plotted in
Fig~\ref{fig10} is obtained with the parameters $SJ=3.23$\,meV,
$SJ_1=SJ_2=-0.15$\,meV, and with $\delta=0.65$ and
$\varepsilon=0.03$ as found earlier. This set of parameters gives
the best fit to the data given the constraints, but is only a
marginal improvement on the case $J_1=J_2=0$.

In the more general case where charge-order leads to orthorhombic
symmetry, $J_1$ and $J_2$ could be different. Let us therefore
consider the case where $J$ and $J_1$ are antiferromagnetic and
$J_2$ is ferromagnetic, consistent with the structure in
Fig.~\ref{fig1}(b). This also creates differences between the magnon
energies at $(0.5,0)$ and at $(1,0)$, and in addition it increases
the energy of the in-plane mode at wavevectors such as $(0.25,0.75)$
and $(0.75,0.25)$. Neither of these effects is observed
experimentally.

The spin-wave model just described succeeds in giving a good
description of the dispersion of the lowest energy branches of the
magnetic excitation spectrum and provides compelling evidence that
the magnetism in La$_{1.5}$Sr$_{0.5}$CoO$_{4}$ is dominated by
180\,deg Co$^{2+}-$Co$^{2+}$ exchange interactions. However, this
model fails in two respects. First, it does not account for the band
of magnetic scattering observed in the range $20-30$\,meV, and
second it cannot predict the correct intensities of the modes
because it ignores orbital angular momentum.

The effective spin--$\frac{1}{2}$ model takes for its basis the
eigenfunctions of $S_x$, where $x$ is the spin quantization
direction. When single-ion anisotropy is strong and the true spin is
greater than $\frac{1}{2}$, as we have here, there can be
significant admixing of the basis states. This means that
excitations to higher single-ion levels can propagate and can be
observed by neutron scattering. Moreover, when the single-ion states
contain a non-negligible orbital component this needs to be included
for an accurate calculation of the neutron cross section, because
the neutron couples to both the spin and the orbital angular
momentum.

To achieve a more complete description of the magnetic spectrum we
use a more realistic linear spin-wave formulation which takes as its
basis the product states $|L_xS_x\rangle$ and determines the
single-ion states self-consistently through the action of the
crystal and exchange fields and the spin--orbit coupling. The method
follows closely the approach described in Ref.\
\onlinecite{Buyers-JPC-1971} in connection with the dispersive
magnetic excitations of antiferromagnetic KCoF$_3$. Similar
calculations have been done for CoO (Refs.\
\onlinecite{Sakurai-PR-1968} and \onlinecite{Tomiyasu-JPSJ-2006})
and for CoF$_2$ (Ref.\ \onlinecite{Allen-PRB-1971}), the latter
though without explicit inclusion of the orbital angular momentum.

The Hamiltonian for the coupled 3$d$ states of Co$^{2+}$ (3$d^7$,
$L=3$, $S=3/2$) in the distorted octahedral ligand field in
La$_{1.5}$Sr$_{0.5}$CoO$_{4}$ is taken to be
\begin{eqnarray}
{\mathcal H}& = & \sum_{\langle jk\rangle}J_{jk}{\bf S}_j \cdot {\bf
S}_k + \sum_j\left[\sum_{l,m} B_l^m O_l^m({\bf L}_j) + \lambda {\bf
L}_j
\cdot {\bf S}_j\right].\nonumber \\[10pt] \label{Eq3}
\end{eqnarray}
The first term is an isotropic Heisenberg superexchange interaction
between true spins $S=\frac{3}{2}$, and the second and third terms
describe the single-ion crystal (ligand) field and spin--orbit
interactions, respectively. The crystal field interaction is assumed
to be the source of magnetic anisotropy, and any anisotropy in the
exchange interactions is neglected. The $O_l^m$ are Stevens operator
equivalents with $B_l^m$ the corresponding crystal-field parameters,
and $\lambda$ is the spin--orbit coupling parameter. We used the
value $\lambda = -18.7$\,meV recently obtained from optical
spectroscopy of CoO.\cite{Kant-PRB-2008} The axially-distorted
octahedral crystal field is represented by the Stevens operators
$O_2^0$, $O_4^0$ and $O_4^4$, where the quantization direction is
taken along the tetragonal axis. We used the point charge
model\cite{Hutchings-SSP-1964} to make a first estimate of the
crystal-field parameters, using the data in Table \ref{table1} for
the average Co--O bond lengths and the results of Ref.
\onlinecite{Zaliznyak-PRB-2001} for the oxygen displacements due to
charge order, which slightly reduces the axial distortion of the
octahedron. This predicted a total splitting of the 3$d^7$ $^4F$
term of about 800\,meV (the small admixture of higher terms was
neglected). The crystal-field splitting in
La$_{1.5}$Sr$_{0.5}$CoO$_{4}$ has not been measured, but optical
measurements of several compounds containing
octahedrally-coordinated CoO$_6$ complexes show a typical splitting
of about 2\,eV.\cite{Abragam,Pratt-PR-1959} This indicates that the
point-charge parameters are about a factor 2 too small, so for an
initial estimate we doubled the point-charge values, giving $B_2^0 =
9.3$\,meV, $B_4^0 = -1.35$\,meV and $B_4^4 = -8.0$\,meV. We note
that with this crystal field the ground state has $XY$-like
anisotropy, as observed experimentally. The ground to first-excited
doublet-doublet splitting is not very sensitive to the $B_4^0$ and
$B_4^4$ parameters, but varies strongly with $B_2^0$. We therefore
adjusted $B_2^0$ to obtain a good fit to the experimental data.

The method for diagonalizing (\ref{Eq3}) is described in the
Appendix.  To compare the many-level spin--orbital model with the
data we fixed $J_1=J_2=0$, since we have already established that
these parameters are very small, and allowed only the parameters $J$
and $B_2^0$ to vary. The best agreement was obtained with $B_2^0 =
13.0$\,meV and $J=1.4$\,meV.  A small anisotropy-field term ${\bf
H}_a\cdot{\bf S}$ with ${\bf H}_a$ directed along the $x$ axis and
of magnitude 0.22\,meV was added to (\ref{Eq3}) to fix the direction
of the ordered moments and to reproduce the observed in-plane spin
gap of $\sim$3\,meV. The dispersion relations and response functions
calculated with these parameters are shown in Fig.~\ref{fig8}. The
calculated dispersion for the low-energy branches is virtually
indistinguishable from that calculated from the effective
spin--$\frac{1}{2}$ model described earlier and agrees very well
with the experimental data. We note that when $J_1=J_2=0$ the
magnetic dispersion is the same for the two twins of the assumed
magnetic structure [Fig.~\ref{fig1}(b)] whose propagation vectors
are related by a 90$^{\circ}$ rotation. This means that the analysis
of the excitation spectrum is not affected by the twinning.

The many-level spin--orbital model allows us to go beyond the lowest
two magnon branches and examine the modes derived from the upper
doublet in the single-ion spectrum (Fig.~\ref{fig9}). The model
predicts two pairs of dispersive bands between 25\,meV and 30\,meV
as shown in the middle panel of Fig.~\ref{fig8}. In the lower panel
of Fig.~\ref{fig8} it can be seen that two out of these four modes
have non-negligible response functions. Interestingly, the mode with
strongest intensity is a longitudinal excitation, labeled $xx$ in
Fig.~\ref{fig8}, and is therefore very different in character to a
conventional spin precession wave.

In Fig.~\ref{fig7} we show neutron scattering intensity maps
calculated with the many-level spin--orbital model alongside the
corresponding experimental data. The presented intensity is
$(k_i/k_f)\,{\rm d}^2\sigma/ {\rm d}\Omega{\rm d}E_{\rm f}$, i.e.\
the scattering cross section in absolute units as defined in
Eq.~\ref{Eq1} multiplied by the factor $k_i/k_f$. The simulations
properly take into account the variation in the scattering vector
$\bf Q$ with position on the detector and with energy, and the
intensity has been averaged over equivalent 90\,deg magnetic
domains. The simulated spectra were convoluted with a Lorentzian
broadening function with a full-width at half maximum of 4\,meV to
take into account the spectrometer resolution ($\simeq 2$\,meV) and
intrinsic broadening, and have been multiplied by a constant scale
factor of 0.3 to obtain quantitative agreement with the experimental
data. We estimate that absorption and self-shielding in the sample
accounts for a factor of about 0.5, and a further reduction of order
10\% may be expected due to zero-point magnetic fluctuations not
included in the model. Therefore, the known corrections account for
a scale factor of about 0.45, which is somewhat larger than the
applied scale factor of 0.3.

Overall, the simulations from the many-level spin--orbital model
provides an excellent description of the observed magnetic
excitation spectrum in the measured energy range.
With a realistic crystal field and only two free parameters in the
Hamiltonian ($J$ and $B_2^0$) plus an overall scale factor the
simulations reproduce the dispersion and intensities of the all the
observed modes in the spectrum extremely well.

\section{Discussion}
\label{Discussion}

The experiments presented here have shown that both the magnetic
order and the magnetic excitation spectrum of charge-ordered
La$_{1.5}$Sr$_{0.5}$CoO$_{4}$ can be understood in terms of a
square, two-sublattice, collinear antiferromagnet, with ordered
moments localized on the Co$^{2+}$ ions and pointing in the CoO$_2$
layers.  The two lowest-frequency branches of the magnetic
excitation spectrum are conventional spin-precession waves, but
higher-frequency dispersive modes are of different character. In
particular, a branch observed at $\sim 25$\,meV is found to
correspond to propagating magnetic fluctuations which are
longitudinal relative to the ordered moment direction.

To arrive at this point, proper account has had to be taken of the
orbital component in the ground state, which is not negligible. For
example, from the self-consistent solution of the mean-field
equations (\ref{App-eq8}) we calculate the spin and orbital
components of the ordered moment to be 2.69\,$\mu_{\rm B}$ and
1.12\,$\mu_{\rm B}$, respectively. The unquenched orbital component
has also been shown to be important for achieving a quantitative
understanding of the bulk susceptibility.\cite{Hollmann-NJP-2008}
The total ordered moment in the model of 3.8\,$\mu_{\rm B}$ is
rather larger than the experimentally-determined ordered moment of
2.9\,$\mu_{\rm B}$ (Ref.~\onlinecite{Zaliznyak-PRL-2000}). It is not
clear why these values should differ, but if confirmed then this
difference could be the reason why the calculated spectrum has a
higher overall intensity than the measured spectrum.

Perhaps the most surprising result to emerge from the analysis of
the spin excitation spectrum is the virtual absence of the 90\,deg
Co$^{2+}$--Co$^{3+}$--Co$^{2+}$ exchange couplings $J_1$ and $J_2$
relative to the 180\,deg coupling $J$. This is dramatically counter
to the simplest estimate $J_1 \sim J_2 \sim 2J$ based on exchange
paths involving a single Co$^{3+}$ bonding
orbital,\cite{Zaliznyak-PRL-2000} and implies that frustration
effects on the magnetic order are very small. One consequence is
that the magnetic structure can be regarded as two interpenetrating,
square-lattice, antiferromagnets with spacing $2a$. In the complete
absence of $J_1$ and $J_2$ the moment directions for the two
interpenetrating antiferromagnets would be unrelated to one another,
but quantum fluctuations are expected to stabilize collinear order
of the global magnetic system via the order-by-disorder
mechanism.\cite{Kim-PRL-1999} A study of the spin wave spectrum of
stripe-ordered La$_{2-x}$Sr$_{x}$NiO$_{4}$ similarly concluded that
the 180\,deg Ni$^{2+}$--Ni$^{2+}$ exchange was significantly larger
than the 90\,deg coupling.\cite{Wu-PRB-2005} Given the importance of
exchange for the stability of spin- and charge-ordered ground states
it would be of interest to seek an understanding the exchange
interactions in these systems in terms of the bonding orbitals
involved.

There remain some details of the magnetic order to be understood. In
Fig.\ \ref{fig5}(a) we reported how the ratio
$I_{110}/I_{1\bar{1}0}$ increases from approximately 0.4 to almost
1.0 in the temperature range $30 - 50$\,K, and in Fig.\ \ref{fig4}
we see that the start of this increase coincides with a kink in the
magnetization below which the field-cooled and zero-field-cooled
magnetization separate. To these observations we add a recent
finding from a muon-spin rotation ($\mu$SR) study
\cite{Giblin-unpublished} that the kink coincides with the
temperature at which magnetic ordering sets in on the muon timescale
($\sim 10^{-6}$\,s). Since $\mu$SR and bulk magnetization probe much
longer timescales than neutron diffraction ($\sim 10^{-12}$\,s) we
can take 31\,K to be the static magnetic ordering temperature.
Between 31\,K and $\sim 60$\,K magnetic Bragg peaks are still
observed by neutron diffraction and so in this temperature range the
magnetic order is not static but fluctuates on a timescale between
$\sim 10^{-6}$ and $\sim 10^{-12}$\,s.

For the assumed collinear magnetic structure there are several
possible ways to explain the behavior of $I_{110}/I_{1\bar{1}0}$.
One is in terms of a canting of the moments.  If the moments point
along the $a$ or $b$ direction then $I_{110}/I_{1\bar{1}0} = 1$, but
if they make an angle greater than 45\,deg to $[110]$ then
$I_{110}/I_{1\bar{1}0} < 1$ [note that we are referring here to the
domain in which the modulation of the magnetic structure is along
$[110]$ --- see Fig.\ \ref{fig1}(b)]. The observed low temperature
value of $I_{110}/I_{1\bar{1}0} = 0.4$ corresponds to a turn-angle
of 12\,deg. According to this model, therefore, the moments point
along the $a$ (or $b$) direction when they first start to order, but
gradually turn towards the $[1\bar{1}0]$ direction (i.e.~away from
the modulation direction) on cooling. By $\sim 30$\,K the moments
have turned through an angle of about 12\,deg, i.e.~in Fig.\
\ref{fig1}(b) $\phi = -12$\,deg or $+102$\,deg, and they remain at
this angle at lower temperatures. Although consistent with the data,
this model requires an explanation for what causes the rotation of
the ordered moments and for why they fix on a canting angle of
12\,deg from the Co--O bond directions at low temperatures.

Another possible interpretation of the $I_{110}/I_{1\bar{1}0}$ data
is in terms of a change in population of different inequivalent spin
domains. For example, if the easy magnetic direction within the
plane were at 45 deg to the Co--O bonds, i.e. $\phi = \pm45$\,deg in
Fig.\ \ref{fig1}(b), then in the ordered phase the moments could in
principle point either parallel or perpendicular to the in-plane
modulation direction of the structure. These longitudinal and
transverse structures are not related by symmetry so their energies
will in general differ. If the energy of the transverse structure
were the lower of the two then the transverse structure would be
present in greater proportion at the temperature of $\sim 30$\,K
when static order sets in, resulting in $I_{110}/I_{1\bar{1}0} < 1$.
With increasing temperature above $\sim 30$\,K thermal fluctuations
would tend to equalize the populations of the two domains,
consistent with the observed increase in $I_{110}/I_{1\bar{1}0}$.
The difference between the field-cooled and zero-field-cooled
magnetization below $30$\,K may be caused by the effect of the
magnetic field favoring one domain over the other.

Keeping these models for the magnetic order in mind let us now
consider the data on the orthogonal components of the in-plane
magnetic fluctuations at 2\,meV and 4\,meV (Section
\ref{section:magnetic_excitations}). Assuming transverse
fluctuations of the moments about an angle $\phi = -12$\,deg with
respect to the Co--O bonds we expect to measure a ratio
$S^{110}/S^{1\bar{1}0} = \cot^2(45^{\circ}+\phi) = 2.4$, whereas
experimentally we find ratios of $1.19 \pm 0.07$ at 2\,meV and $1.00
\pm 0.05$ at 4\,meV corresponding to an angle of at most 1 or 2
\,deg away from the Co--O bond direction. Because the scattering is
proportional to the square of the orthogonal components of the
moments the model with unequal domain populations predicts the same
ratio $S^{110}/S^{1\bar{1}0} = 2.4$. Both models, therefore, are
apparently at odds with the data.

However, the elastic and inelastic scattering data can be reconciled
through consideration of the particular form of the magnetic order
which, as discussed above, can be regarded as two interpenetrating,
square-lattice antiferromagnets with spacing $2a$. In isolation,
each of these antiferromagnets would have a magnetic dispersion with
four-fold symmetry so that, for example, there would be equivalent
minima in the dispersion at each of the in-plane wavevectors
$\pm(0.25,0.25)$, and $\pm(0.25,-0.25)$. Weak coupling between the
two antiferromagnets locks them into a single structure with two
domains each having a two-fold pattern of magnetic Bragg peaks, but
the four-fold symmetry will remain in the excitation spectrum above
a crossover energy related to the coupling between the two
antiferromagnets. This has two consequences for the data here. One
is that after averaging the excitation spectra from the two
wavevector domains the in-plane fluctuations are isotropic above the
crossover energy. Therefore the observation of
$S^{110}/S^{1\bar{1}0} \simeq 1$ at 2\,meV and 4\,meV suggests that
the crossover energy is below 2\,meV and is further evidence that
the magnetic order comprises two weakly-coupled, interpenetrating
antiferromagnets. The other consequence is that the magnon
dispersion surface above the crossover energy is the same for both
wavevector domains and corresponds to a square-lattice collinear
antiferromagnet with spacing $2a$. This means that domain-averaging
and the fact that we do not have a unique model for the magnetic
order has no bearing on the interpretation of the excitation
spectra.

Finally, we note that we have not found any evidence for magnetic
degrees of freedom associated with the Co$^{3+}$ site in the
magnetic spectrum probed here up to $\sim 50$\,meV. However, we
cannot rule out the possibility of a small Van-Vleck moment on the
Co$^{3+}$ sites induced by coupling to the Co$^{2+}$ moments. If
this were the case then such a coupling might be able to cause a
spin-canting transition, and this might be another possible
explanation for the $I_{110}/I_{1\bar{1}0}$ data. In the absence of
experimental evidence to test this possibility we do not attempt to
speculate further, but since other spin--charge ordered systems
exhibit similar changes in $I_{110}/I_{1\bar{1}0}$ with temperature
to that found
here\cite{Freeman-PRB-2002,Lee-PRB-2001,Freeman-PRB-2004,Giblin-PRB-2008}
it would be of interest to examine this behavior more closely.


\section{Conclusions}

We have gained a rather complete understanding of the nature of the
magnetic excitations in what is a text-book charge-ordered,
two-dimensional antiferromagnet. The magnetic order is stabilized
essentially by a single exchange interaction acting along a
straight-line path between the charge-ordered Co$^{2+}$ sites. We
find no evidence for active magnetic degrees of freedom on the
Co$^{3+}$ sites. Open questions include what is precise nature of
the magnetic order and how to explain the exchange in terms of the
bonding.

\section*{ACKNOWLEDGMENTS}

A.T.B. is grateful to the Laboratory for Neutron Scattering at the
Paul Scherrer Institute for hospitality and support during an
extended visit in 2009. We thank Roger Cowley and Giniyat Khaliullin
for valuable discussions, and Sean Giblin for communicating the
$\mu$SR results prior to publication. This work was supported by the
Engineering \& Physical Sciences Research Council of Great Britain.

\section*{APPENDIX: GENERALIZED LINEAR SPIN-WAVE THEORY}

\subsection{Effective spin--$\frac{1}{2}$ model for ground state doublet}

Quantization of Eq.~(\ref{Eq2}) by means of the Holstein--Primakoff
transformation, followed by diagonalization of the resulting
Hamiltonian by the standard method,\cite{White-PR-1965} leads to two
non-degenerate branches with in-plane dispersion
\begin{equation}
\hbar\omega_{1,2}({\bf Q}) = \sqrt{(A_{\bf Q}\pm B_{\bf Q})^2-
D^2_{\bf Q}},\label{App-eq2}
\end{equation}
where $\omega_1$ ($\omega_2$) corresponds to the upper (lower) sign,
and
\begin{eqnarray}
A_{\bf Q} & = & 2S\{2J(1+\varepsilon) + J_{1}-J_{2} +J_{2}\cos[{\bf
Q}\cdot ({\bf a}-{\bf b})] \}
\nonumber\\[5pt] B_{\bf Q} &
= & S\{J\delta\cos(2{\bf Q}\cdot{\bf
a})+J\delta\cos(2{\bf Q}\cdot{\bf b})\},\nonumber\\[5pt] D_{\bf Q} & = & 2S\{J(1-\delta/2)[\cos(2{\bf Q}\cdot{\bf
a})+\cos(2{\bf Q}\cdot{\bf b})] \nonumber\\[2pt]
& & \hspace{15pt} + J_1\cos[{\bf Q}\cdot({\bf a}+{\bf b})]\}.
\label{App-eq3}
\end{eqnarray}
With the spins aligned along $x$, only the transverse correlations
$yy$ and $zz$ contribute to the linear spin-wave cross section, Eq.~
(\ref{Eq1}). The corresponding response functions (per
La$_{1.5}$Sr$_{0.5}$CoO$_{4}$ formula unit) for magnon creation are
given by
\begin{eqnarray} S^{yy}({\bf Q},\omega) & =
& \frac{g_y^2S}{4}\frac{A_{\bf Q}+B_{\bf Q}-D_{\bf
Q}}{\hbar\omega_{1}({\bf Q})}
\{n(\omega)+1\}\delta[\omega-\omega_{1}({\bf Q})], \nonumber\\[5pt] S^{zz}({\bf Q},\omega) & =
& \frac{g_z^2S}{4}\frac{A_{\bf Q}-B_{\bf Q}-D_{\bf
Q}}{\hbar\omega_{2}({\bf Q})}
\{n(\omega)+1\}\delta[\omega-\omega_{2}({\bf Q})],\nonumber\\[5pt]
\label{App-eq5}
\end{eqnarray}
where $g_y$ and $g_z$ are in-plane and out-of-plane $g$-factors for
the effective spin--$\frac{1}{2}$ ground-state doublet of the
Co$^{2+}$ ion and $n(\omega)$ is the boson occupation number. We
neglect the corrections which are sometimes applied to the spin-wave
dispersion and response functions to account for zero-point
fluctuations.

\subsection{Many-level spin--orbital model}

We diagonalize (\ref{Eq3}) in two steps. First, we diagonalize the
single-ion terms, i.e.\ the crystal field and spin--orbit terms,
plus the molecular field part of the exchange energy. This produces
a set of self-consistent single-ion energy levels and wavefunctions
for each site. The second step is to write the residual part of the
exchange interaction in terms of pseudo-boson raising and lowering
operators for the single-ion states, retaining terms up to quadratic
order. The resulting Hamiltonian is bilinear in the pseudo-boson
operators and can be diagonalized by the standard procedure.

To be specific, let us assume the antiferromagnetic order in
La$_{1.5}$Sr$_{0.5}$CoO$_{4}$ to be composed of two sublattices $A$
and $B$, with the $A$-sublattice moments along $+x$ and the
$B$-sublattice moments along $-x$. The Hamiltonian is the sum of a
single-ion part ${\mathcal H}_1$ and a two-ion part ${\mathcal
H}_2$. The single-ion Hamiltonian for one unit cell is given by
\begin{equation}
{\mathcal H}_1 = {\mathcal H}_1^A + {\mathcal H}_1^B,
\label{App-eq7}
\end{equation}
where
\begin{eqnarray}
{\mathcal H}_1^A & = & {\mathcal H}_{\rm cf}^A + {\mathcal H}_{\rm
so}^A + {\bf S}^A \cdot
{\bf H}_{\rm mf}^A,\nonumber\\
[5pt] {\mathcal H}_1^B & = & {\mathcal H}_{\rm cf}^B + {\mathcal
H}_{\rm so}^B + {\bf S}^B \cdot {\bf H}_{\rm
mf}^B\hspace{5pt}.\label{App-eq8}
\end{eqnarray}
In the first of these equations ${\mathcal H}_{\rm cf}^A$,
${\mathcal H}_{\rm so}^A$ and ${\bf H}_{\rm mf}^A$ are the crystal
field, spin--orbit and molecular field interactions for the $A$
site. The latter is given by
\begin{equation}
{\bf H}_{\rm mf}^A = \langle{\bf
S}^B\rangle\sum_{\{\Delta_B\}}J_{\Delta_B}+\langle{\bf
S}^A\rangle\sum_{\{\Delta_A\}}J_{\Delta_A}.\label{App-eq9}
\end{equation}
$A$ and $B$ are interchanged for the $B$ site terms in
(\ref{App-eq8}). $\Delta_A$ and $\Delta_B$ represent the
displacements from an $A$ site to other $A$ and $B$ sites, and
$J_{\Delta_A}$ and $J_{\Delta_B}$ are the corresponding exchange
parameters. In practice, we restrict the model to
nearest-neighboring sites, so the summations over the $B$- and
$A$-site neighbors in Eq.\ (\ref{App-eq9}) amount to $4J+2J_1$ and
$2J_2$, respectively (see Fig.\ \ref{fig1}). Note that $\langle{\bf
S}^B\rangle = -\langle{\bf S}^A\rangle$.

The mean-field equations (\ref{App-eq8}) are solved
self-consistently by an iterative process until the values of
$\langle{\bf S}^A\rangle$ and $-\langle{\bf S}^B\rangle$ converge to
an acceptable level of precision (in our case one part in $10^6$).
This gives the set of single-ion energy levels $\epsilon_n^A =
\epsilon_n^B = \epsilon_n$, where $n$ takes values from 0 (ground
state) to $(2L+1)\times(2S+1)-1 = 27$. From the corresponding
single-ion wavefunctions $|n\rangle$, the matrix elements for spin
\begin{equation}
{\bf S}_{n'n}=\langle n' |{\bf S}|n\rangle\label{App-eq10}
\end{equation}
and for the total magnetic moment
\begin{equation}
{\bf M}_{n'n}=-\langle n' |{\bf L}+2{\bf S}|n\rangle\label{App-eq11}
\end{equation}
can be calculated for the $A$ and the $B$ sites.

We now consider the two-ion part of the Hamiltonian which describes
the residual exchange interactions. By symmetry, the interaction
energy is the same for the two sites, so we can write it as twice
the energy for the $A$ site,
\begin{eqnarray}
{\mathcal H}_2 & = & \sum_{\{\Delta\}}J_{\Delta}{\bf S}^A\cdot{\bf
S}^{\Delta} - 2{\bf S}^A \cdot {\bf H}_{\rm
mf}^A\nonumber\\[5pt] & = & \sum_{\{\Delta\}}J_{\Delta}\,({\bf S}^A-\langle{\bf S}^A\rangle)
\cdot({\bf S}^{\Delta}-\langle{\bf S}^{\Delta}\rangle) -\langle{\bf
S}^A\rangle\cdot{\bf H}_{\rm mf}^A \hspace{2pt}.\nonumber\\
\label{App-eq12}
\end{eqnarray}
The summation is over all sites connected to the $A$ site by
non-zero exchange interactions. The molecular field part of the
exchange interaction is subtracted because this is included in the
single-ion terms --- see Eqs.\ (\ref{App-eq8}).

To quantize the Hamiltonian we introduce pseudo-boson raising and
lowering operators which convert the ground state into the excited
states, and vice-versa. For the $A$ site,
\begin{eqnarray}
a_n^{\dag}|0\rangle  =  |n\rangle \hspace{15pt} {\rm and}
\hspace{15pt} a_n|n\rangle  = |0\rangle\hspace{2pt}.\label{App-eq13}
\end{eqnarray}
Operators $b_n^{\dag}$ and $b_n$ are defined similarly for the $B$
site. If the temperature is sufficiently low that the equilibrium
population of the ground state is close to one then to a good
approximation these operators satisfy the Bose commutation
relations\cite{Grover-PR-1965}
\begin{equation}
[a_n,a_{n'}^{\dag}] = [b_n,b_{n'}^{\dag}] =
\delta_{nn'}\hspace{2pt}. \label{App-eq14}
\end{equation}
Operators on different sites commute. The single-ion Hamiltonian can
now be written as
\begin{equation}
{\mathcal H}_1 = \sum_{n>0} \epsilon_n(a_n^{\dag}a_n +
b_n^{\dag}b_n)\hspace{2pt}.\label{App-eq15}
\end{equation}

For the two-ion Hamiltonian we start with the following identity for
the spin operator on the $A$ site,
\begin{eqnarray}
{\bf S} & = & {\bf S}_{00} + \sum_{n>0}[\,{\bf
S}_{n0}a_n^{\dag}+{\bf S}_{0n}a_n + ({\bf S}_{nn}-{\bf
S}_{00})a_n^{\dag}a_n\,]\nonumber\\[5pt]
& & \hspace{18pt}+\sum\sum_{\hspace{-10pt} n\neq n' >0}{\bf
S}_{n'n}a_{n'}^{\dag}a_n\hspace{2pt}. \label{App-eq16}
\end{eqnarray}
At low temperatures ${\bf S}_{00} \simeq \langle {\bf S}\rangle$,
and we need only retain the linear terms in the operators if we are
to neglect higher-than-quadratic terms in the Hamiltonian. With
these approximations Eq.~(\ref{App-eq16}) simplifies to
\begin{equation}
{\bf S}-\langle {\bf S}\rangle = \sum_{n>0} {\bf
S}_{n0}a_n^{\dag}+{\bf S}_{0n}a_n\hspace{2pt},\label{App-eq17}
\end{equation}
and the two-ion Hamiltonian (\ref{App-eq12}) becomes
\begin{widetext}
\begin{eqnarray}
{\mathcal H}_2 = -\langle{\bf S}^A\rangle\cdot{\bf H}_{\rm mf}^A
\hspace{5pt}+&&\hspace{-5pt}\sum_{n,n'>0}\sum_{\Delta_A}J_{\Delta_A}[\,({\bf
S}_{n0}^A\cdot{\bf S}_{0n'}^A)a_n^{\dag}a_{n',\Delta_A}+({\bf
S}_{n0}^A\cdot{\bf
S}_{n'0}^A)a_n^{\dag}a_{n',\Delta_A}^{\dag}+ {\rm h.c.}\,] \nonumber\\
+&&\hspace{-5pt}\sum_{n,n'>0}\sum_{\Delta_B}J_{\Delta_B}[\,({\bf
S}_{n0}^A\cdot{\bf S}_{0n'}^B)a_n^{\dag}b_{n',\Delta_B}+({\bf
S}_{n0}^A\cdot{\bf S}_{n'0}^B)a_n^{\dag}b_{n',\Delta_B}^{\dag}+ {\rm
h.c.}\,]\hspace{2pt}.\label{App-eq18}
\end{eqnarray}
\end{widetext}
The Fourier transform operators are defined by
\begin{eqnarray}
a_{{\bf m}+{\bm \Delta}} & = & \frac{1}{\sqrt{N}}\sum_{\bf
Q}\exp[i{\bf
Q}\cdot({\bf m}+{\bm \Delta})]\,a_{\bf Q}\hspace{2pt},\nonumber\\
\hspace{15pt}a_{{\bf m}+{\bm \Delta}}^{\dag} & = &
\frac{1}{\sqrt{N}}\sum_{\bf Q}\exp[-i{\bf Q}\cdot({\bf m}+{\bm
\Delta})]\,a_{\bf Q}^{\dag}\hspace{2pt}, \label{App-eq19}
\end{eqnarray}
where $N$ is the total number of $A$ sites (or $B$ sites). In
Eq.~(\ref{App-eq19}) we explicitly show the position vectors for the
operators: $\bf m$ is position vector for the $A$ site in the
$m^{\rm th}$ unit cell, so for example $a_{{\bf m}+{\bm
\Delta}}^{\dag}$ is the raising operator for the $A$ site that is
displaced from $\bf m$ by $\bm \Delta$. The definitions of the
Fourier transform operators for the $B$ site are the same as those
for the $A$ site except $b$ replaces $a$. After substitution of the
expressions in Eqs.~(\ref{App-eq19}) into (\ref{App-eq15}) and
(\ref{App-eq18}) and summation over $\bf m$ the total Hamiltonian
can be written in the form
\begin{eqnarray}
{\mathcal H} = {\mathcal H}_0 + \frac{1}{2}\sum_{\bf
Q}\sum_{n,n'>0}{\bf X}^{\dag}_{{n},{\bf Q}}\textsf{H}_{{nn'},{\bf
Q}}{\bf X}_{{n'},{\bf Q}}\;, \label{App-eq20}
\end{eqnarray}
where ${\mathcal H}_0$ contains the constant terms, ${\bf
X}^{\dag}_{{n},{\bf Q}}$ is the row matrix $(a_{n,{\bf
Q}}^{\dag},b_{n,{\bf Q}}^{\dag},a_{n,-{\bf Q}},b_{n,-{\bf Q}})$ for
the excited level $n$, ${\bf X}_{{n'},{\bf Q}}$ is the column matrix
containing the Hermitian adjoint operators, and
\begin{eqnarray}
 \textsf{H}_{{nn'},{\bf Q}} =
 \left( \begin{array}{l l l l}
                   A_{nn',{\bf Q}} & B_{nn',{\bf Q}} & C_{nn',{\bf Q}} & D_{nn',{\bf Q}}\\
                   B^{\ast}_{n'n,{\bf Q}} & A_{nn',{\bf Q}} & D_{n'n,{-\bf Q}} & C_{nn',{\bf Q}}\\
                   C^{\ast}_{nn',{\bf Q}} & D^{\ast}_{nn',{-\bf Q}} & A_{n'n,{\bf Q}} & B^{\ast}_{nn',{-\bf Q}}\\
                   D^{\ast}_{n'n,{\bf Q}} & C^{\ast}_{nn',{\bf Q}} & B_{nn',{-\bf Q}} & A_{n'n,{\bf Q}}
                  \end{array}\right) \, .\nonumber\\
\label{App-eq21}
\end{eqnarray}
The coefficients in the matrix are
\begin{eqnarray}
A_{nn',{\bf Q}} & = & \epsilon_n\delta_{nn'}+({\bf
S}_{n0}^A\cdot{\bf S}_{0n'}^A)
\gamma^A_{\bf Q}\,,\nonumber\\
B_{nn',{\bf Q}} & = & ({\bf S}_{n0}^A\cdot{\bf S}_{0n'}^B)
\gamma^B_{\bf Q}\,,\nonumber\\
C_{nn',{\bf Q}} & = & ({\bf S}_{n0}^A\cdot{\bf S}_{n'0}^A)
\gamma^A_{\bf Q}\,,\nonumber\\
D_{nn',{\bf Q}} & = & ({\bf S}_{n0}^A\cdot{\bf S}_{n'0}^B)
\gamma^B_{\bf Q}\,,\label{App-eq22}
\end{eqnarray}
where
\begin{eqnarray}
\gamma^A_{\bf Q} & = & \sum_{\Delta_A} J_{\Delta_A}\exp(i{\bf Q}\cdot{{\bm \Delta}_A})\,,\nonumber\\
\gamma^B_{\bf Q} & = & \sum_{\Delta_B} J_{\Delta_B}\exp(i{\bf
Q}\cdot{{\bm \Delta}_B})\,.\label{App-eq23}
\end{eqnarray}
In general, $\gamma^A_{-\bf Q}=\gamma^A_{\bf Q}$ because all $A$
sites are equivalent on the magnetic lattice. For the present
system,
\begin{eqnarray}
\gamma^A_{\bf Q} & = & 2J_2\cos[{\bf Q}\cdot({\bf a}-{\bf
b})]\,,\nonumber\\[5pt]
\gamma^B_{\bf Q} & = & 2J\cos(2{\bf Q}\cdot{\bf a})+2J\cos(2{\bf
Q}\cdot{\bf b})\nonumber\\ & & + 2J_1\cos[{\bf Q}\cdot({\bf a}+{\bf
b})]\,.\label{App-eq24}
\end{eqnarray}
The Hamiltonian (\ref{App-eq20}) can now be diagonalized by the
standard method.\cite{White-PR-1965} There are a total of
$(2L+1)\times(2S+1)-1 = 27$ single-ion excited levels for Co$^{2+}$
in La$_{1.5}$Sr$_{0.5}$CoO$_{4}$ giving a total of 54 distinct modes
in the magnetic spectrum, two modes for each single-ion excited
level $n$. In our case we diagonalized the full Hamiltonian which is
represented by a $108 \times 108$ matrix (since each mode appears
twice in the Hamiltonian).

To evaluate the neutron scattering cross section we employ the
general form for the response function that takes into account
orbital as well as spin magnetization. For the creation of one
magnon in the $|n_j\rangle$ mode ($j=1,2$) from the fully ordered
ground state via $\alpha\alpha$ correlations the response function
(per La$_{1.5}$Sr$_{0.5}$CoO$_{4}$ formula unit) is given by
\begin{eqnarray}
S^{\alpha\alpha}({\bf Q},\omega) = \frac{1}{4}\left|\,\langle
n_j|M^{\alpha}({\bf
Q})\,|0\rangle\right|^2\,\delta[\omega-\omega_{n_j}({\bf
Q})]\,.\nonumber\\
\label{App-eq25}
\end{eqnarray}
By replacing $\bf S$ by $\bf M$ in Eq.~(\ref{App-eq17}) and summing
over the $A$ and $B$ sites we obtain the following expression for
the operator representing the Fourier transform of the
magnetization,
\begin{eqnarray}
M^{\alpha}({\bf Q}) =  \sum_{n>0} &&
[\,(M^{\alpha}_{n0})^Aa^{\dag}_{{n},{\bf
Q}}+(M^{\alpha}_{0n})^Aa_{{n},{-\bf Q}}\nonumber\\
 && +(M^{\alpha}_{n0})^Bb^{\dag}_{{n},{\bf
Q}}+(M^{\alpha}_{0n})^Bb_{{n},{-\bf Q}}\,] \,.\nonumber\\
\label{App-eq26}
\end{eqnarray}
The $a^{\dag}_{{n},{\bf Q}}$, $a_{{n},{-\bf Q}}$, etc., operators
are expressed as a linear combination of creation and destruction
operators for the magnon modes via the Bogoliubov transformation
matrix. From the coefficients of the creation operator for the
$|n_j\rangle$ mode one can calculate the matrix element in
(\ref{App-eq26}), and hence the response function
Eq.~(\ref{App-eq25}) and scattering cross section Eq.~(\ref{Eq1})
for magnon creation in this mode. The resulting scattering intensity
and response functions for the lowest six modes of
La$_{1.5}$Sr$_{0.5}$CoO$_{4}$ are shown in Figs.~\ref{fig7} and
\ref{fig8}.


\begin{references}

\bibitem{Tranquada-Nature-1995}
J. M. Tranquada,  B. J. Sternleib, J. D. Axe, Y. Nakamura, and S.
Uchida, Nature (London) {\bf 375}, 561 (1995).

\bibitem{Chen-PRL-1993}
C. H. Chen, S-W. Cheong, and A. S. Cooper, Phys. Rev. Lett. {\bf
71}, 2461 (1993).

\bibitem{Tranquada-PRL-1994}
J. M. Tranquada, D. J. Buttrey, V. Sachan, and J. E. Lorenzo, Phys.
Rev. Lett. {\bf 73}, 1003 (1994).

\bibitem{Yamada-PhysicaC-1994}
K. Yamada, T. Omata, K. Nakajima, Y. Endoh, and S. Hosoya, Physica C
{\bf 221}, 355 (1994).

\bibitem{Cwik-PRL-2009}
M. Cwik, M. Benomar, T. Finger, Y. Sidis, D. Senff, M. Reuther, T.
Lorenz, and M. Braden,  Phys. Rev. Lett. {\bf 102}, 057201 (2009).

\bibitem{Mn-half-doped}
Y. Moritomo, Y. Tomioka, A. Asamitsu, Y. Tokura, and Y. Matsui,
Phys. Rev. B {\bf 51}, 3297 (1995); B. J. Sternlieb, J. P. Hill, U.
C. Wildgruber, G. M. Luke, B. Nachumi, Y. Moritomo, and Y. Tokura,
Phys. Rev. Lett. {\bf 76}, 2169 (1996); Y. Murakami, H. Kawada, H.
Kawata, M. Tanaka, T. Arima, Y. Moritomo, and Y. Tokura, Phys. Rev.
Lett. {\bf 80}, 1932 (1998).

\bibitem{Zaliznyak-PRL-2000}
I. A. Zaliznyak, J. P. Hill, J. M. Tranquada, R. Erwin, and Y.
Moritomo, Phys. Rev. Lett. {\bf 85}, 4353 (2000).

\bibitem{No-Mn-charge-order}
P. Mahadevan, K. Terakura, and D. D. Sarma, Phys. Rev. Lett. {\bf
87}, 066404 (2001); J. Wang, W. Zhang, and D. Y. Xing, J. Phys.:
Condens. Matter {\bf 14}, 4659 (2002).

\bibitem{Kajimoto-PRB-2003}
R. Kajimoto, K. Ishizaka, H. Yoshizawa, and Y. Tokura, Phys. Rev.
B {\bf 67}, 014511 (2003).

\bibitem{Freeman-PRB-2002}
P. G. Freeman, A. T. Boothroyd, D. Prabhakaran, D. Gonz\'{a}lez, and
M. Enderle, Phys. Rev. B {\bf 66}, 212405 (2002).

\bibitem{Moritomo-PRB-1997}
Y. Moritomo, K. Higashi, K. Matsuda, and A. Nakamura, Phys. Rev. B
{\bf 55}, R14725 (1997).

\bibitem{Yamada-PRB-1989}
K. Yamada, M. Matsuda, Y. Endoh, B. Keimer, R. J. Birgeneau, S.
Onodera, J. Mizusaki, T. Matsuura, and G. Shirane, Phys. Rev. B {\bf
39}, 2336 (1989).

\bibitem{Zaliznyak-PRB-2001}
I. A. Zaliznyak, J. M. Tranquada, R. Erwin, and Y. Moritomo, Phys.
Rev. B {\bf 64}, 195117 (2001).


\bibitem{Hollmann-NJP-2008}
N. Hollmann, M. W. Haverkort, M. Cwik, M. Benomar, M. Reuther, A.
Tanaka, and T. Lorenz, New J. Phys. {\bf 10}, 023018 (2008).

\bibitem{Chang-PRL-2009}
C. F. Chang, Z. Hu, H. Wu, T. Burnus, N. Hollmann, M. Benomar, T.
Lorenz, A. Tanaka, H.-J. Lin, H. H. Hsieh, C. T. Chen, and L. H.
Tjeng, Phys. Rev. Lett. {\bf 102}, 116401 (2009).

\bibitem{Raccah-PR-1967}
P. M. Raccah and J. B. Goodenough, Phys. Rev. {\bf 155}, 932 (1967).

\bibitem{Zaliznyak-JAP-2004}
I. A. Zaliznyak, J. M. Tranquada, G. Gu, R. W. Erwin, Y. Moritomo,
J. Appl. Phys. {\bf 95}, 7369 (2004).

\bibitem{Savici-PRB-2007}
A. T. Savici, I. A. Zaliznyak, G. D. Gu, and R. Erwin, Phys. Rev. B
{\bf 75}, 184443 (2007).

\bibitem{Helme-PhysicaC-2004}
L. M. Helme, A. T. Boothroyd, D. Prabhakaran, F. R. Wondre, C. D.
Frost, and J. Kulda, Physica B {\bf 350}, e273, (2004).

\bibitem{Horigane-JMMM-2007}
K. Horigane, K. Yamada, H. Hiraka, and J. Akimitsu, J. Magn. Magn.
Mater. {\bf 310}, 774 (2007).

\bibitem{Horigane-JPSJ-2007}
K. Horigane, H. Hiraka, T. Uchida, K. Yamada, and J. Akimitsu, J.
Phys. Soc. Jpn. {\bf 76}, 114715 (2007).

\bibitem{Freeman-PRB-2005}
P. G. Freeman, A. T. Boothroyd, D. Prabhakaran, C. D. Frost, M.
Enderle, and A. Hiess, Phys. Rev. B {\bf 71}, 174412 (2005).

\bibitem{Senff-PRL-2006}
D. Senff, F. Kr\"{u}ger, S. Scheidl, M. Benomar, Y. Sidis, F.
Demmel, and M. Braden, Phys. Rev. Lett. {\bf 96}, 257201 (2006).

\bibitem{Prabhak-JCG-2005}
D. Prabhakaran, A. T. Boothroyd, F. R. Wondre, and T. J. Prior, J.
Cryst. Growth {\bf 275}, e827 (2005).

\bibitem{Moon-PR-1969}
R. M. Moon and T. Riste and W. C. Koehler, Phys. Rev. {\bf 181}, 920
(1969).

\bibitem{GSAS}
A. C. Larson and R. B. Von Dreele, ``General Structure Analysis
System (GSAS)", Los Alamos National Laboratory Report LAUR 86-748
(2000).

\bibitem{LeToquin-PhysicaB-2004}
R. Le Toquin, W. Paulus, A. Cousson, G. Dhalenne, and A.
Revcolevschi, Physica B {\bf 350}, e269 (2004).

\bibitem{Helme-DPhil-2006}
L. M. Helme, DPhil Thesis, University of Oxford (2006). Available
from \url{http://xray.physics.ox.ac.uk/Boothroyd}.

\bibitem{Lee-PRB-2001}
S. -H. Lee, S. -W. Cheong, K. Yamada, and C. F. Majkrzak, Phys. Rev.
B {\bf 63}, 060405(R) (2001).

\bibitem{Freeman-PRB-2004}
P. G. Freeman, A. T. Boothroyd, D. Prabhakaran, M. Enderle, and C.
Niedermayer, Phys. Rev. B {\bf 70}, 024413 (2004).

\bibitem{Giblin-PRB-2008}
S. R. Giblin, P. G. Freeman, K. Hradil, D. Prabhakaran, and A. T.
Boothroyd, Phys. Rev. B {\bf 78}, 184423 (2008).

\bibitem{Squires}
G. L. Squires, {\it Introduction to the theory of thermal neutron
scattering} (Dover Publications, Inc., 1996).

\bibitem{Buyers-JPC-1971}
W. J. L. Buyers, T. M. Holden, E. C. Svensson, R. A. Cowley, and M.
T. Hutchings, J. Phys. C {\bf 4}, 2139 (1971).

\bibitem{Sakurai-PR-1968}
J. Sakurai, W. J. L. Buyers, R. A. Cowley, G. Dolling, Phys. Rev.
{\bf 167}, 510 (1968).

\bibitem{Tomiyasu-JPSJ-2006}
K. Tomiyasu and S. Itoh, J. Phys. Soc. Jpn., {\bf 75},084708 (2006).

\bibitem{Allen-PRB-1971}
S. J. Allen, Jr. and H. J. Guggenheim, Phys. Rev. B {\bf 4}, 950
(1971).

\bibitem{Kant-PRB-2008}
Ch. Kant, T. Rudolf, F. Schrettle, F. Mayr, J. Deisenhofer, P.
Lunkenheimer, M. V. Eremin, and A. Loidl, Phys. Rev. B {\bf 78},
245103 (2008).

\bibitem{Hutchings-SSP-1964}
M. T. Hutchings, Solid State Phys. {\bf 16}, 227 (1964).

\bibitem{Abragam}
A. Abragam and B. Bleaney, {\it Electron Paramagnetic Resonance of
Transition Ions} (Oxford University Press, 1970).

\bibitem{Pratt-PR-1959}
G. W. Pratt and R. Coelho, Phys. Rev. {\bf 116}, 281 (1959).



\bibitem{Kim-PRL-1999}
Y. J. Kim, A. Aharony, R. J. Birgeneau, F. C. Chou, O.
Entin-Wohlman, R. W. Erwin, M. Greven, A. B. Harris, M. A. Kastner,
I. Ya. Korenblit, Y. S. Lee, and G. Shirane, Phys. Rev. Lett. {\bf
83}, 852 (1999).

\bibitem{Wu-PRB-2005}
H. Woo, A. T. Boothroyd, K. Nakajima, T. G. Perring, C. D. Frost, P.
G. Freeman, D. Prabhakaran, K. Yamada and J. M. Tranquada, Phys.
Rev. B {\bf 72}, 064437 (2005).

\bibitem{Giblin-unpublished}
S. R. Giblin and P. G. Freeman, unpublished.

\bibitem{White-PR-1965}
R. M. White, M. Sparks, and I. Ortenburger, Phys. Rev. {\bf 139},
A450 (1965).

\bibitem{Grover-PR-1965}
B. Grover, Phys. Rev. {\bf 140}, A1944 (1965).



\end{references}
\end{document}